\documentclass[preprint,eqsecnum,aps,nofootinbib]{revtex4}
\usepackage{amsfonts,amsmath,amssymb,amsthm}
\usepackage{latexsym}
\usepackage{bbm,bm}
\usepackage{graphicx}

\newcommand{\tr}{\mathrm{tr}}

\newcommand{\ket}[1]{\lvert #1 \rangle}
\newcommand{\bra}[1]{\langle #1 \lvert}
\newcommand{\beq}{\begin{equation}}
\newcommand{\eeq}{\end{equation}}
\newcommand{\beqs}{\begin{eqnarray}}
\newcommand{\eeqs}{\end{eqnarray}}
\begin{document}

\title{Thermal Entanglement Phase Transition in Coupled Harmonic Oscillators with Arbitrary Time-Dependent Frequencies}

\author{DaeKil Park$^{1,2}$}

\affiliation{$^1$Department of Electronic Engineering, Kyungnam University, Changwon
                 631-701, Korea    \\
             $^2$Department of Physics, Kyungnam University, Changwon
                  631-701, Korea    
                      }

\begin{abstract}
We derive explicitly the thermal state of the two coupled harmonic oscillator system when the spring and coupling constants are 
arbitrarily time-dependent. In particular, we focus on the case of sudden change of frequencies. In this case we compute purity 
function,  R\'{e}nyi and von Neumann entropies, and mutual information analytically and examine their temperature-dependence. We also discuss on the 
thermal entanglement phase transition by making use of the negativity-like quantity. Our calculation shows that the critical temperature $T_c$ increases with 
increasing the difference between the initial and final frequencies. In this way we can protect the entanglement against the external temperature by introducing  large difference of initial and final frequencies. 
\end{abstract}

\maketitle
\section{Introduction}
Entanglement\cite{schrodinger-35,text,horodecki09} is a key physical resource in quantum information processing. For example, 
it plays crucial role in  quantum teleportation\cite{teleportation},
superdense coding\cite{superdense}, quantum cloning\cite{clon}, quantum cryptography\cite{cryptography,cryptography2}, quantum
metrology\cite{metro17}, and quantum computer\cite{qcomputer,qcreview}. In particular, physical realization of quantum cryptography and quantum computer seems to be accomplished in the near future\footnote{see Ref. \cite{white} and web page
https://www.computing.co.uk/ctg/news/3065541/european-union-reveals-test-projects-for-first-tranche-of-eur1bn-quantum-computing-fund.}. 

Although entanglement is highly useful property of quantum state, it is normally fragile when quantum system interacts with its surroundings. 
Interaction with the environments makes the given quantum system undergo decoherence\cite{zurek03} and as a result, it loses its quantum properties.
Thus, decoherence significantly changes the quantum entanglement. Sometimes entanglement exhibits an exponential decay in time by successive halves. 
Sometimes, however, entanglement sudden death (ESD) occurs when the entangled multipartite quantum system is
embedded in Markovian environments\cite{markovian,yu05-1,yu06-1,yu09-1,almeida07,park-16}. This means that the entanglement is completely disentangled at finite times. 

Most typical surrounding is external temperature. At finite temperature quantum mechanics the external temperature is introduced via imaginary time at 
zero temperature quantum mechanics. Thus, the exponential decay or ESD-like phenomenon can occur in external temperature. If external temperature induces 
the ESD-like phenomenon in temperature, this means there exists a critical temperature $T_c$, below or above which the entanglement of a system is 
nonzero or completely zero. We will call this phenomenon thermal entanglement phase transition (TEPT) between nonzero entanglement phase and 
zero entanglement phase. The TEPT and the critical temperature $T_c$ were explored\cite{park-19} recently by making use of concurrence\cite{form2, form3}
in anisotropic Heisenberg $X Y Z$ spin model with Dzyaloshinskii-Moriya interaction\cite{dzya58,mori60}.

The purpose of this paper is to study on the TEPT phenomenon in continuous variable system. Most simple continuous variable system seems to be two-coupled harmonic 
oscillator system. In this reason we will choose this system to explore the TEPT when the  spring constant $k_0$ and coupling constant $J$ are arbitrarily 
time-dependent. Another reason we choose this system is because of the fact that the thermal state of this system is Gaussian. It is known that the 
Peres-Horodecki positive partial transposition (PPT) criterion\cite{peres96,horodecki96,horodecki97} provides a necessary and sufficient condition for separability of 
Gaussian continuous variable states\cite{duan-2000,simon-2000}. Thus, the temperature-dependence of entanglement can be roughly deduced 
by considering the negativity-like quantity\cite{vidal01}. What we are interested in is to examine how the arbitrarily time-dependent parameters affect the 
critical temperature. In particular, we focus in this paper on the sudden quenched model, where the system parameters abruptly change at $t=0$. 

The paper is organized as follows. In section II we derive the thermal state of single harmonic oscillator system when the frequency is arbitrarily 
time-dependent. We focus on the case of sudden quenched model (SQM). For SQM we derive the purity function and von Neumann entropy
of the thermal state analytically.  In section III we derive explicitly the thermal state of two coupled harmonic oscillator system when  the spring constant $k_0$
and coupling constant $J$ are arbitrarily time-dependent. In section IV  we compute the purity function,  R\'{e}nyi and von Neumann entropies, and mutual information analytically for the thermal state of two coupled harmonic oscillator system in the case of SQM. It is shown that the thermal state is less mixed with 
increasing the difference between initial and final frequencies at the given external temperature. The mutual information shows that  the common information parties $A$ and $B$ share does not completely vanish even in the infinity temperature limit. In section V the TEPT is discussed for the case of SQM 
by making use of the negativity-like quantity. It is shown that the critical temperature $T_c$ increases with increasing the frequency difference. Thus, using SQM with large difference of initial and final frequencies it seems to be possible to protect entanglement against external temperature. In section VI a brief conclusion is given. In appendix A the eigenvalue equation for the thermal state of the coupled harmonic oscillator system is explicitly solved.

\section{Thermal State for single harmonic oscillator with arbitrary time-dependent frequency}
Let us consider a single harmonic oscillator with time-dependent frequency, whose Hamiltonian is 
\begin{equation}
\label{revise-hamil-1}
H_1 = \frac{1}{2} p^2 + \frac{1}{2} \omega^2 (t) x^2.
\end{equation}
Then, the action functional of this system is given by 
\begin{equation}
\label{action}
S[x] = \int_0^t \left[ \frac{1}{2} \dot{x}^2 - \frac{1}{2} \omega^2 (t) x^2 \right].
\end{equation}
Usually Kernel for any quantum system can be derived by computing the path-integral\cite{feynman}
\begin{equation}
\label{path-integral}
K[x', x : t] = \int_{(0,x)}^{(t,x')} {\cal D}x e^{i S[x]}.
\end{equation}
Although the path-integral with constant frequency can be computed\cite{feynman,kleinert}, it does not seem to be simple 
matter to compute the path-integral explicitly when $\omega$ is arbitrary time-dependent. However, it is possible to derive the Kernel without computing the 
path-integral if one uses the Schr\"{o}dinger description of Kernel
\begin{equation}
\label{finaleq}
K[{\bm x'}, t_2 : {\bm x}, t_1] = \sum_n \psi_n \left( {\bm x'}, t_2 \right) \psi_n^* \left( {\bm x}, t_1 \right)
\end{equation}
where $n$ represents all possible quantum numbers and $\psi_n \left({\bm x}, t \right)$ is linearly-independent solution of time-dependent 
Schr\"{o}dinger equation (TDSE). 

The TDSE of our system was exactly solved in Ref. \cite{lewis68,lohe09}. The linearly independent solutions $\psi_n (x, t) \hspace{.1cm} (n=0, 1, \cdots)$ are expressed in a form
\begin{equation}
\label{TDSE-1}
\psi_n (x, t) = e^{-i E_n \tau(t)} e^{\frac{i}{2} \left( \frac{\dot{b}}{b} \right) x^2} \phi_n \left( \frac{x}{b} \right)
\end{equation}
where
\begin{eqnarray}
\label{TDSE-2}
&& E_n = \left( n + \frac{1}{2} \right) \omega(0),    \hspace{1.0cm}  \tau (t) = \int_0^t \frac{d s}{b^2 (s)}       \\   \nonumber
&&\phi_n (x) = \frac{1}{\sqrt{2^n n!}} \left( \frac{ \omega (0)} {\pi b^2} \right)^{1/4} H_n \left(\sqrt{\omega (0)} x \right) e^{-\frac{\omega (0)}{2} x^2 }.
\end{eqnarray}
In Eq. (\ref{TDSE-2}) $H_n (z)$ is $n^{th}$-order Hermite polynomial and $b(t)$ satisfies the Ermakov equation
\begin{equation}
\label{ermakov-1}
\ddot{b} + \omega^2 (t) b = \frac{\omega^2 (0)}{b^3}
\end{equation}
with $b(0) = 1$ and $\dot{b} (0) = 0$. 

Solutions of the Ermakov equation were discussed in Ref. \cite{pinney50}. If $\omega(t)$ is time-independent, $b(t)$ is simply one. 
If $\omega (t)$ is instantly changed as
\begin{eqnarray}
\label{instant-1}
\omega (t) = \left\{                \begin{array}{cc}
                                               \omega_0  & \hspace{1.0cm}  t = 0   \\
                                               \omega & \hspace{1.0cm}  t > 0,
                                               \end{array}            \right.
\end{eqnarray}
then $b(t)$ becomes
\begin{equation}
\label{scale-1}
b(t) = \sqrt{ \frac{\omega^2 - \omega_0^2}{2 \omega^2} \cos (2 \omega t) +  \frac{\omega^2 + \omega_0^2}{2 \omega^2}}.
\end{equation}
For more general time-dependent case the Ermakov equation should be solved numerically.

Inserting Eq. (\ref{TDSE-1}) into Eq. (\ref{finaleq}) and using 
\begin{equation}
\label{formula-1}
\sum_{n=0}^{\infty} \frac{t^n}{n!} H_n (x) H_n (y) = (1 - 4 t^2)^{-1/2} \exp \left[ \frac{4 t x y - 4 t^2 (x^2 + y^2)}{1 - 4 t^2} \right],
\end{equation}
Kernel for this system becomes
\begin{equation}
\label{kernel-single}
K[x', x: t]  = e^{\frac{i}{2} \left( \frac{\dot{b}}{b} \right) x'^2} \frac{(\omega_0 \omega')^{1/4}}{\sqrt{2 \pi i \sin \Gamma(t)}}
\exp \left[\frac{i}{2 \sin \Gamma(t)} \left\{ (\omega_0 x^2 + \omega' x'^2) \cos \Gamma(t) - 2 \sqrt{\omega_0 \omega'} x x' \right\} \right]
\end{equation}
where
\begin{equation}
\label{boso-1}
\omega_0 = \omega (t = 0), \hspace{1.0cm} \omega'(t) = \frac{\omega_0}{b^2 (t)}, \hspace{1.0cm} \Gamma (t) = \int_0^t \omega'(s) d s.
\end{equation}

For time-independent case $b(t) = 1$, $\omega_0 = \omega' = \omega$, and $\Gamma (t) = \omega t$. Then, the Kernel in Eq. (\ref{kernel-single}) reduces to 
usual well-known harmonic oscillator Kernel
\begin{equation}
\label{usual1}
K[x', x : t] = \sqrt{\frac{\omega}{2 \pi i \sin \omega t}} \exp \left[ \frac{i \omega}{2 \sin \omega t} 
\left\{ (x^2 + x'^2) \cos \omega t - 2 x x' \right\} \right].
\end{equation}
It is remarkable to note that the $x \leftrightarrow x'$ symmetry in Eq. (\ref{usual1}) is broken in Eq. (\ref{kernel-single}) due to the time-dependence of 
frequency. In fact, it is manifest due to the fact that the system parameters at $t=0$ are different from those at $t > 0$. 

From now on in this section we consider only the case of SQM given in Eq. (\ref{instant-1}). In this case the Kernel becomes 
\begin{equation}
\label{sudden-1}
K[x', x : t] = e^{- \left( \frac{i (\omega^2 - \omega_0^2)}{4 \omega b^2} \sin 2 \omega t \right) x'^2} \sqrt{\frac{\omega_0}{2 \pi i b \sin \Gamma(t)}}
\exp \left[ \frac{i \omega_0}{2 \sin \Gamma(t)} \left\{ \left(x^2 + \frac{x'^2}{b^2} \right) \cos \Gamma(t) - \frac{2 x x'}{b} \right\} \right]
\end{equation}
where $b(t)$ is given in Eq. (\ref{scale-1}) and $\Gamma (t)$ becomes 
\begin{equation}
\label{sudden-2}
\Gamma (t) = \tan^{-1} \left(\frac{\omega_0}{\omega} \tan \omega t \right) 
= \frac{1}{2 i} \ln \frac{\omega + i \omega_0 \tan \omega t}{\omega - i \omega_0 \tan \omega t}.
\end{equation}

In quantum mechanics the inverse temperature $\beta = 1 / k_B T$ is introduced as a Euclidean time $\beta = i t$ (see Ch. $10$ of Ref.\cite{feynman}), 
where $k_B$ is a Boltzmann constant. Then, the thermal density matrix is defined as
\begin{equation}
\label{thermal-1}
\rho_T [x', x : \beta] = \frac{1}{{\cal Z} (\beta)} G[x', x : \beta]
\end{equation}
where $\beta = 1 / k_B T$, $G[x', x : \beta] = K[x', x : -i \beta]$, and ${\cal Z} (\beta) = \mbox{tr} G[x', x : \beta]$ is a partition function.
Throughout this paper we use $k_B = 1$ for convenience. For SQM case $b(t)$ and $\Gamma (t)$ are changed into $b (\beta)$ and 
$\Gamma (\beta)$, whose explicit expressions are 
\begin{equation}
\label{sudden-3}
b(\beta) = \sqrt{ \frac{\omega^2 - \omega_0^2}{2 \omega^2} \cosh (2 \omega \beta) +  \frac{\omega^2 + \omega_0^2}{2 \omega^2}}, \hspace{1.0cm}
\Gamma (\beta) = -i \Gamma_0 (\beta)
\end{equation}
where\footnote{In fact, one can show that $b (\beta)$ in Eq. (\ref{sudden-3}) is a solution of $\frac{d^2 b}{d \beta^2} - \omega^2 b = - \frac{\omega_0^2}{b^3}$.}
\begin{equation}
\label{sudden-4}
\Gamma_0 (\beta) = \frac{1}{2}  \ln \left(\frac{\omega +  \omega_0 \tanh \omega \beta}{\omega -  \omega_0 \tanh \omega \beta}\right).
\end{equation}
Then, the partition function of this system becomes
\begin{equation}
\label{partition-1}
{\cal Z} (\beta) = \sqrt{\frac{\omega_0}{2 \pi b \sinh \Gamma_0}} \sqrt{\frac{\pi}{a_{0, -}}}
\end{equation}
where
\begin{equation}
\label{partition-2}
a_{0, \pm} (\beta) = A_0 (\beta) + \frac{\omega_0}{2 \sinh \Gamma_0} \left[ \left( 1 + \frac{1}{b^2} \right) \cosh \Gamma_0 \pm \frac{2}{b} \right]
\end{equation}
with
\begin{equation}
\label{partition-3}
A_0 (\beta) = \frac{\omega^2 - \omega_0^2}{4 \omega b^2} \sinh (2 \omega \beta).
\end{equation}
Using the partition function one can derive the thermal density matrix in a form  
\begin{equation}
\label{thermal-2}
\rho_0 [x', x : \beta] = \sqrt{\frac{a_{0, -}}{\pi}} e^{-A_0 (\beta) x'^2} 
\exp\left[ - \frac{\omega_0}{2 \sinh \Gamma_0} \left\{ \left( x^2 + \frac{x'^2}{b^2} \right) \cosh \Gamma_0 - \frac{2 x x'}{b} \right\} \right].
\end{equation}

The thermal density matrix is in general mixed state. In order to explore how much it is mixed we first compute the purity function 
$P_0 (\beta) = \mbox{tr} \left( \rho_0 \right)^2$. If it is one, this means that $\rho_0$ is pure state. If it is zero, this means $\rho_0$ is completely mixed state. 
If $0 < P_0 (\beta) < 1$, this means that $\rho_0$ is partially mixed state. It is not difficult to show that the purity function of this system is
\begin{equation}
\label{purity-1}
P_0 (\beta) \equiv \int dx dx' \rho_0 [ x', x : \beta] \rho_0 [ x, x' : \beta]  = \sqrt{\frac{a_{0,-}}{a_{0,+}}}.
\end{equation}

Another quantity we want to compute is a von Neumann entropy $S[\rho_0]$ of $\rho_0$. If $\rho_0$ is pure state, $S[\rho_0]$ is zero. If its 
mixedness increases, $S[\rho_0]$ also increases from zero and eventually goes to infinity for completely mixed state in this continuum case. In order to compute 
the von Neumann entropy we should solve the eigenvalue equation 
\begin{equation}
\label{von-1}
\int d x \rho_0 [x', x : \beta] f_n (x) = \lambda_n (\beta) f_n (x').
\end{equation}
One can show that the eigenvalue equation
\begin{equation}
\label{von-2}
\int d x \left( A e^{-a_1 x^2 - a_2 x'^2 + 2 b x x'} \right) f_n (x) = \lambda_n f_n (x')
\end{equation}
can be solved, and the eigenfunction and corresponding eigenvalue are 
\begin{eqnarray}
\label{von-3}
&&f_n (x) = {\cal C}_n^{-1}  H_n (\sqrt{\epsilon_0} x) e^{-\frac{\alpha_0}{2} x^2}                        \\    \nonumber
&&\lambda_n = A \sqrt{\frac{2 \pi}{(a_1 + a_2) + \epsilon_0}} \left[ \frac{(a_1 + a_2) - \epsilon_0}{(a_1 + a_2) + \epsilon_0} \right]^{n/2}
\end{eqnarray}
where $\epsilon_0 = \sqrt{(a_1 + a_2)^2 - 4 b^2}$ and $\alpha_0 = \epsilon_0 - (a_1 - a_2)$.
By making use of integral formula\cite{integral}
\begin{eqnarray}
\label{integral2-1}
&&\int_{-\infty}^{\infty} e^{-(x - y)^2} H_{m} (c x) H_n (c x) d x      \\     \nonumber
&&= \sqrt{\pi} \sum_{k=0}^{\min (m, n)} 2^k k! \left(  \begin{array}{c} m  \\  k  \end{array}  \right) 
\left(  \begin{array}{c} n  \\  k  \end{array}  \right) (1 - c^2)^{\frac{m+n}{2} - k} H_{m+n - 2 k} \left( \frac{c y}{\sqrt{1 - c^2}} \right)
\end{eqnarray}
and various properties of Gamma function\cite{handbook}, the normalization constant ${\cal C}_n$ can be written in a form 
\begin{equation}
\label{normalization2-1}
{\cal C}_n^2 = \frac{1}{\sqrt{\alpha_0}} \sum_{k=0}^n 2^{2 n - k} \left( \frac{\epsilon_0}{\alpha_0} - 1 \right)^{n-k}
\frac{\Gamma^2 (n+1) \Gamma \left(n - k + \frac{1}{2} \right)}{\Gamma (k+1) \Gamma^2 (n - k + 1)}.
\end{equation}
If $a_1 = a_2$, $\alpha_0 = \epsilon_0$, which makes nonzero in $k$-summation of Eq. (\ref{normalization2-1}) only when $k = n$. Then, 
${\cal C}_n$ becomes usual harmonic oscillator normalization constant 
\begin{equation}
\label{normalization2-2}
{\cal C}_n ^{-1} = \frac{1}{\sqrt{2^n n!}} \left( \frac{\epsilon_0}{\pi} \right)^{1 / 4}.
\end{equation}

Using Eqs. (\ref{von-2}) and (\ref{von-3}) the eigenvalue of 
Eq. (\ref{von-1}) becomes
\begin{equation}
\label{von-4}
\lambda_n (\beta) = (1 - \xi_0) \xi_0^n
\end{equation}
where 
\begin{equation}
\label{von-5}
\xi_0 = \frac{\sqrt{a_{0,+}} - \sqrt{a_{0,-}}}{\sqrt{a_{0,+}} + \sqrt{a_{0,-}}} = \frac{1 - P_0 (\beta)} {1 + P_0 (\beta)}.
\end{equation}
Thus, the spectral decomposition of $\rho_0$ can be written as 
\begin{equation}
\label{single-spectral}
\rho_0 [x', x : \beta] = \sum_n 
\lambda_n (\beta) f_n (x': \beta) f_n^* (x: \beta),
\end{equation}
where $f_n (x: \beta)$ is given by Eq. (\ref{von-3}) with $\epsilon_0 = \sqrt{a_{0,+} a_{0,-}}$ and 
$\alpha_0 = \epsilon_0 + A_0 (\beta) - \frac{\omega_0 \cosh \Gamma_0}{2 \sinh \Gamma_0} \left( 1 - \frac{1}{b^2} \right).$
Eq. (\ref{von-4}) implies $\sum_n \lambda_n (\beta) = 1$, which is consistent with $\tr \rho_0 = 1$. Then the von Neumann entropy of $\rho_0$ becomes
\begin{equation}
\label{von-6}
S[\rho_0] \equiv - \sum_n \lambda_n (\beta) \ln \lambda_n (\beta) = - \ln (1 - \xi_0) - \frac{\xi_0}{1 - \xi_0} \ln \xi_0.
\end{equation}

%%%%%%%%%%%%%%%%%%%%%%%%%%%%%%%%%%%%%%%%%%%%%%%%%%%%%%%%%
\begin{figure}[ht!]
\begin{center}
\includegraphics[height=5.0cm]{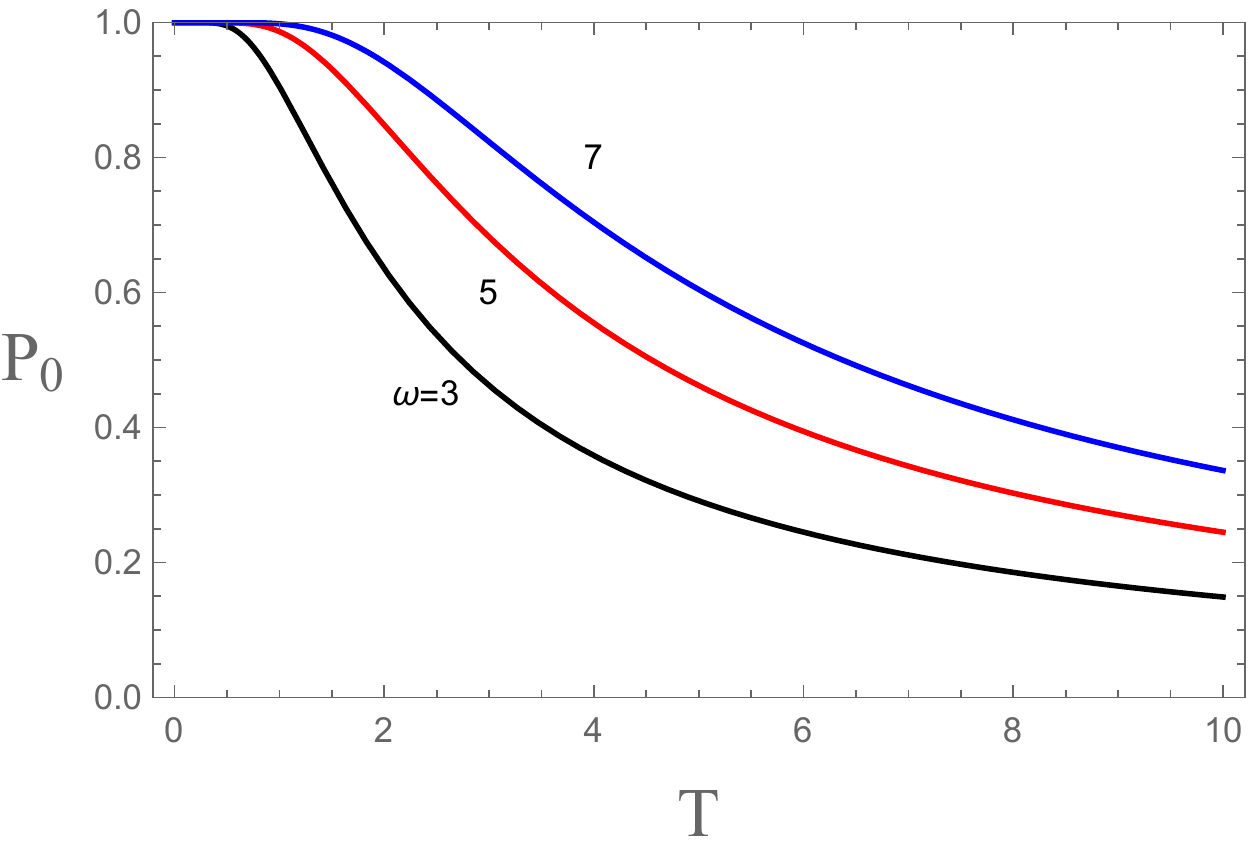} \hspace{0.5cm}
\includegraphics[height=5.0cm]{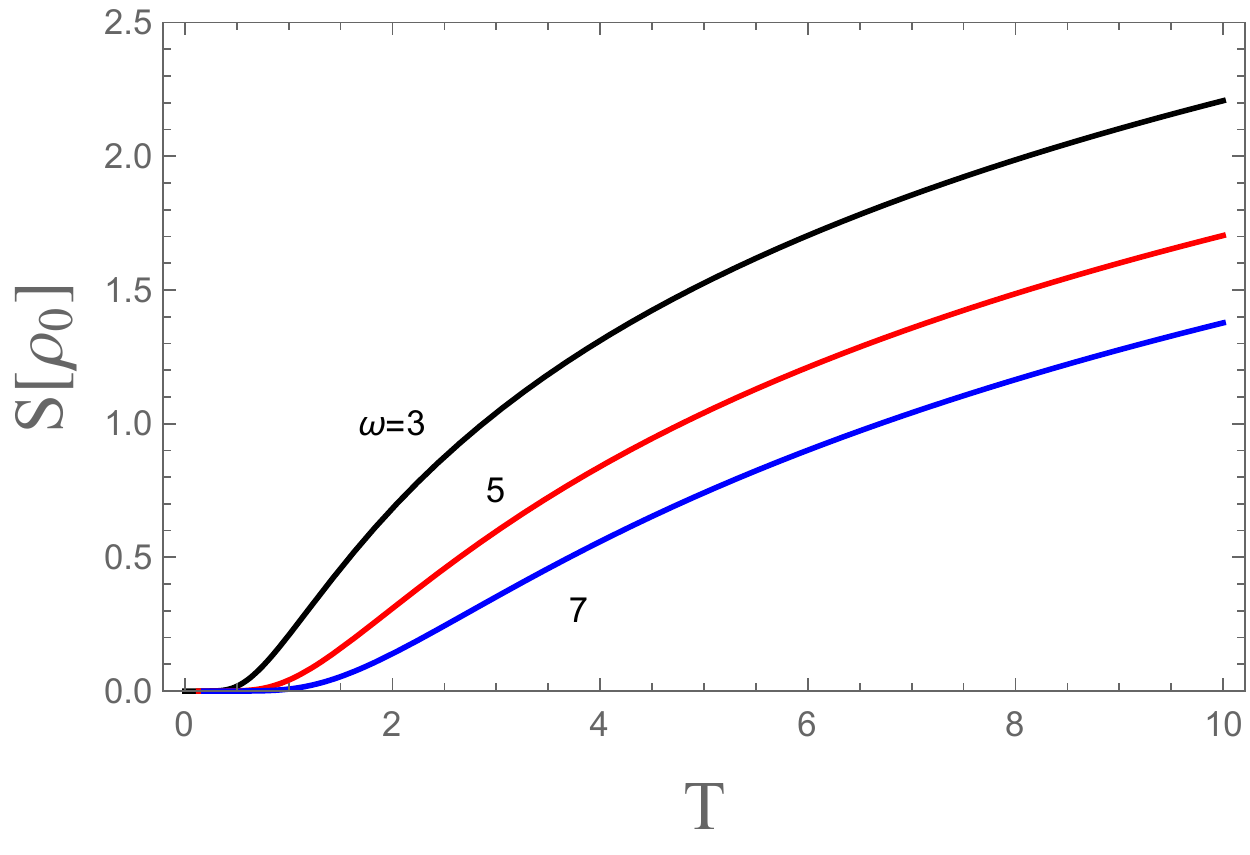}

\caption[fig1]{(Color online) The $T$-dependence of (a) purity function and (b) von Neumann entropy when $\omega = 3$ (black line), 
$5$ (red line), 
$7$ (blue line) with $\omega_0 = 3$. Both figures show that the mixedness of the thermal state (\ref{thermal-2}) decreases with increasing $|\omega - \omega_0|$ at the given temperature.  All figures show that $\rho_0$ becomes more and more mixed with increasing the external temperature.
 }
\end{center}
\end{figure}
%%%%%%%%%%%%%%%%%%%%%%%%%%%%%%%%%%%%%%%%%%%%%%%%%%%%%%%%%%%

For constant frequency, {\it i.e.} $\omega = \omega_0$, $A_0 = 0$, $a_{0,+} = \omega \coth \frac{\omega \beta}{2}$, 
$a_{0,-} = \omega \tanh \frac{\omega \beta}{2}$, and $\xi_0 = e^{-\omega \beta}$. For the case of SQM $A_0$ and $a_{0,\pm}$ become
larger than those in constant frequency case in the entire range of temperature. As a result, $P_0 (\beta)$ and $\xi_0$ become larger and smaller compared to the
constant frequency case. Since $-\ln (1 - x) - \frac{x}{1 - x} \ln x$ is monotonically increasing function in the range $0 \leq x \leq 1$, this fact decreases the 
von Neumann entropy.

The temperature dependence of the purity function and von Neumann entropy is plotted in Fig. 1(a) and Fig. 1(b) when $\omega = 3$ (black line), $5$ (red line), 
$7$ (blue line) with $\omega_0 = 3$. All figures show that $\rho_0$ becomes more and more mixed with increasing the external temperature. Both figures also show that 
$\rho_0$ becomes less mixed with increasing $|\omega - \omega_0|$ at the given temperature. Thus, we can use SQM model to protect the purity against the external temperature.

 \section{Thermal State for two coupled harmonic oscillators with arbitrary time-dependent frequencies} 
In this section we will derive the thermal state for two coupled harmonic oscillator system, whose Hamiltonian is 
\begin{equation}
\label{revise-hamil-2}
H_2 = \frac{1}{2} \left(p_1^2 + p_2^2 \right) + V(x_1, x_2).
\end{equation}
We choose the potential $V(x_1, x_2)$ as a quadratic function with arbitrary time-dependent spring and coupling parameters. 
The explicit expression of the potential is chosen in a form
\begin{equation}
\label{revise-potential-1}
V(x_1, x_2) = \frac{1}{2} \left[ k_0 (t) (x_1^2 + x_2^2) + J(t) (x_1 - x_2)^2 \right].
\end{equation}
Then, the action functional of this system is  
\begin{equation}
\label{action2}
S[x_1, x_2] = \int_0^t dt \left[ \frac{1}{2} \left( \dot{x}_1^2 + \dot{x}_2^2 \right) - V(x_1, x_2) \right].
\end{equation}

It is easy to show that the potential is diagonalized by introducing $y_1 = \frac{1}{\sqrt{2}} (x_1 + x_2)$ and  $y_2 = \frac{1}{\sqrt{2}} (x_1 - x_2)$ . 
In terms of new canonical variables the action becomes that of two non-interacting harmonic oscillators in a form 
\begin{equation}
\label{action3}
S[x_1, x_2] = \int_0^t dt \left[ \frac{1}{2} \left( \dot{y}_1^2 + \dot{y}_2^2 \right) + \frac{1}{2} \left\{\omega_1^2 (t) y_1^2 + \omega_2^2 (t) y_2^2 \right\} \right]
\end{equation}
where $\omega_1 (t) = \sqrt{k_0 (t)}$ and $\omega_2 (t) = \sqrt{k_0 (t) + 2 J (t)}$. Thus, the Kernel for this system is 
\begin{eqnarray}
\label{kernel2-1}
&&K[x_1', x_2':x_1, x_2: t] =                                                                                                               \\   \nonumber
&&\prod_{j=1}^2 \left[ e^{\frac{i}{2} \left( \frac{\dot{b}_j}{b_j} \right) y_j'^2} \frac{\left(\omega_{j,0} \omega_j'\right)^{1/4}}{\sqrt{2 \pi i \sin \Gamma_j (t)}}  \exp \left[\frac{i}{2 \sin \Gamma_j (t)} \left\{ (\omega_{j,0} y_j^2 + \omega_j' y_j'^2) \cos \Gamma_j (t) - 2 \sqrt{\omega_{j,0} \omega_j'} 
    y_j y_j'  \right\}                                     \right]                                                                                                                     \right]
\end{eqnarray}
where $\omega_{j,0} = \omega_j (t = 0)$, $\omega_j' = \frac{\omega_{j,0}}{b_j^2 (t)}$, and $\Gamma_j (t) = \int_0^t \omega_j' (s) ds$.  Of course, 
$b_1 (t)$ and $b_2 (t)$ satisfy the Ermakov equation 
\begin{equation}
\label{ermakov2-1}
\ddot{b}_j + \omega_j^2 (t) b_j = \frac{\omega_{j,0}}{b_j^3}    \hspace{1.0cm} (j = 1, 2)
\end{equation}
with $\dot{b}_j (0) = 0$ and $b_j (0) = 1$. Then the thermal density matrix of this system is given by 
\begin{equation}
\label{thermal2-1}
\rho_T[x_1', x_2':x_1, x_2:\beta] = \frac{1}{{\cal Z} (\beta)} K[x_1',x_2':x_1,x_2:-i \beta]
\end{equation}
where $\beta = 1 / k_B T$ and ${\cal Z} (\beta) = \mbox{tr}  K[x_1',x_2':x_1,x_2:-i \beta]$. 

In this paper we will examine only the case of SQM. More general time-dependent cases will be explored elsewhere. If spring and 
coupling constants are abruptly changed as 
\begin{eqnarray}
\label{instant2-1}
k_0 = \left \{          \begin{array}{cc}
                    k_{0,i}   &   \hspace{1.0cm} t = 0    \\
                    k_{0,f}   &   \hspace{1.0cm} t > 0
                             \end{array}                                    \right. 
\hspace{2.0cm}
J = \left \{          \begin{array}{cc}
                    J_{i}   &   \hspace{1.0cm} t = 0    \\
                    J_{f}   &   \hspace{1.0cm} t > 0,
                             \end{array}                                     \right.
\end{eqnarray}
$\omega_1$ and $\omega_2$ become
\begin{eqnarray}
\label{instant2-2}
\omega_1 =  \left \{          \begin{array}{cc}
               \omega_{1,0} \equiv \omega_{1,i} = \sqrt{k_{0,i}}  &   \hspace{.3cm} t = 0    \\
               \omega_{1,f} = \sqrt{k_{0,f}}   &  \hspace{.3cm}   t > 0,
                                          \end{array}                                                 \right.
\hspace{0.7cm}
\omega_2 =  \left \{          \begin{array}{cc}
               \omega_{2,0} \equiv \omega_{2,i} = \sqrt{k_{0,i} + 2 J_i}  &   \hspace{.3cm} t = 0    \\
               \omega_{2,f} = \sqrt{k_{0,f} + 2 J_f}   &  \hspace{.3cm}   t > 0.
                                          \end{array}                                                 \right.
\end{eqnarray}
Then, the thermal density matrix of this system is given by 
\begin{eqnarray}
\label{thermal2-2}
&&\rho_T[x_1', x_2':x_1,x_2:\beta]                               \\   \nonumber
&&=\prod_{j=1}^2 \sqrt{\frac{a_{j,-}}{\pi}} \exp \left[ -A_j y_j'^2 - \frac{\omega_{j,i}}{2 \sinh \Gamma_{E,j}} 
\left\{ \left( y_j^2 + \frac{y_j'^2}{b_j^2} \right) \cosh \Gamma_{E,j} - \frac{2 y_j y_j'}{b_j} \right\}                   \right]
\end{eqnarray}
where\footnote{The subscript $E$ in $\Gamma_{E,j}$ stands for ``Euclidean''. This subscript is attached to stress the point that the inverse temperature $\beta$ is introduced as a Euclidean time.}
\begin{eqnarray}
\label{sudden2-1}
&& b_j = \sqrt{\frac{\omega_{j,f}^2 - \omega_{j,i}^2}{2 \omega_{j,f}^2} \cosh (2 \omega_{j,f} \beta) + \frac{\omega_{j,f}^2 + \omega_{j,i}^2}
{2 \omega_{j,f}^2}}, 
\hspace{.2cm} \Gamma_{E,j} = \frac{1}{2} \ln \frac{\omega_{j,f} + \omega_{j,i} \tanh (\omega_{j,f} \beta)}{\omega_{j,f} - \omega_{j,i} \tanh (\omega_{j,f} \beta)}                                                                                                                    \\    \nonumber
&&A_j = \frac{\omega_{j,f}^2 - \omega_{j,i}^2}{2 \omega_{j,f} b_j^2} \sinh (2 \omega_{j,f} \beta),   \hspace{.2cm}
a_{j,\pm} = A_j + \frac{\omega_{j,i}}{2 \sinh \Gamma_{E,j}} \left[ \left(1 + \frac{1}{b_j^2} \right) \cosh \Gamma_{E,j} \pm \frac{2}{b_j} \right]
\end{eqnarray}
with $j=1,2$. For the limit of $\omega_{j,i} = \omega_{j,f} \equiv \omega_j$, we have $A_j = 0$, $b_j = 1$, $\Gamma_{E,j} = \omega_j \beta$, 
$a_{j,+} = \omega_j \coth (\omega_j \beta / 2)$, and $a_{j,-} = \omega_j \tanh (\omega_j \beta / 2)$. In terms of $x_j$-coordinates the thermal state 
reduces to 
\begin{eqnarray}
\label{thermal2-3}
&&\rho_T[x_1', x_2':x_1,x_2:\beta]  = \frac{\sqrt{a_{1,-}a_{2,-}}}{\pi} \exp \bigg[ -\alpha_1 (x_1'^2 + x_2'^2) - \alpha_2 (x_1^2 + x_2^2) 
                                                                                                                                                                                                 \\   \nonumber
&&\hspace{2.0cm}+ 2 \alpha_3 x_1' x_2' + 2 \alpha_4 x_1 x_2 + 2 \alpha_5 (x_1 x_1' + x_2 x_2') + 2 \alpha_6 (x_1 x_2' + x_2 x_1')  \bigg]
\end{eqnarray}
where
\begin{eqnarray}
\label{instant2-3}
&&\alpha_1 = \sum_{j=1}^{2} \left[ \frac{A_j}{2} + \frac{\omega_{j,i} \cosh \Gamma_{E,j}}{4 b_j^2 \sinh \Gamma_{E,j}} \right],
\hspace{.5cm} \alpha_2 = \sum_{j=1}^2 \frac{\omega_{j,i} \cosh \Gamma_{E,j}}{4 \sinh \Gamma_{E,j}}    \\   \nonumber
&& \alpha_3 = \sum_{j=1}^2 (-1)^j  \left[ \frac{A_j}{2} + \frac{\omega_{j,i} \cosh \Gamma_{E,j}}{4 b_j^2 \sinh \Gamma_{E,j}} \right],
\hspace{.5cm} \alpha_4 = \sum_{j=1}^2 (-1)^j  \frac{\omega_{j,i} \cosh \Gamma_{E,j}}{4 \sinh \Gamma_{E,j}}          \\    \nonumber
&&\hspace{1.0cm} \alpha_5 = \sum_{j=1}^2 \frac{\omega_{j,i}}{4 b_j \sinh \Gamma_{E,j}},     \hspace{.5cm}
\alpha_6 = \sum_{j=1}^2 (-1)^{j-1}  \frac{\omega_{j,i}}{4 b_j \sinh \Gamma_{E,j}}.
\end{eqnarray}
It is worthwhile noting that $\alpha_j$ satisfy
\begin{eqnarray}
\label{instant2-4}
&&\hspace{3.0cm}\alpha_1 + \alpha_2 = \frac{(a_{1,+} + a_{1,-}) + (a_{2,+} + a_{2,-})}{4}            \\   \nonumber
&&\hspace{3.0cm}\alpha_3 + \alpha_4 = -  \frac{(a_{1,+} + a_{1,-}) - (a_{2,+} + a_{2,-})}{4}         \\    \nonumber
&&\alpha_5 = \frac{1}{8} \left[ (a_{1,+} - a_{1,-}) + (a_{2,+} - a_{2,-})  \right],       \hspace{.5cm}
\alpha_6 =  \frac{1}{8} \left[ (a_{1,+} - a_{1,-}) - (a_{2,+} - a_{2,-})  \right].
\end{eqnarray}
Using Eq. (\ref{instant2-4}) it is straightforward to show $\mbox{tr} \rho_T = 1$. In next section we compute several quantum information quantities analytically, 
which measure how much $\rho_T$ is mixed. 

\section{Various Quantities of Thermal State: Case of SQM}

In this section we will compute purity function, R\'{e}nyi and von Neumann entropies, and mutual information of $\rho_T[x_1', x_2':x_1, x_2:\beta]$ given 
in Eq. (\ref{thermal2-2}) or equivalently Eq. (\ref{thermal2-3}). As a by-product we derive the spectral decomposition of  $\rho_T[x_1', x_2':x_1, x_2:\beta]$.

\subsection{Purity function}

The purity function is defined as 
\begin{equation}
\label{purity3-1}
P(\beta) = \mbox{tr} \rho_T^2 \equiv 
\int dx_1' dx_2' dx_1 dx_2  \rho_T[x_1', x_2':x_1, x_2:\beta]  \rho_T[x_1, x_2:x_1', x_2':\beta].
\end{equation}
If $P(\beta) = 1$ or $0$, this means that $\rho_T$ is pure or completely mixed state. Direct calculation shows 
\begin{equation}
\label{purity3-2}
P(\beta) = \sqrt{\frac{a_{1,-} a_{2,-}}{a_{1,+} a_{2,+}}}.
\end{equation}
For the case of constant frequencies $\omega_{j,i} = \omega_{j,f} \equiv \omega_j$ it reduces to 
\begin{equation}
\label{purity3-3}
P(\beta) = \tanh \frac{\omega_1 \beta}{2} \tanh \frac{\omega_2 \beta}{2}.
\end{equation}

%%%%%%%%%%%%%%%%%%%%%%%%%%%%%%%%%%%%%%%%%%%%%%%%%%%%%%%%%
\begin{figure}[ht!]
\begin{center}
\includegraphics[height=5.0cm]{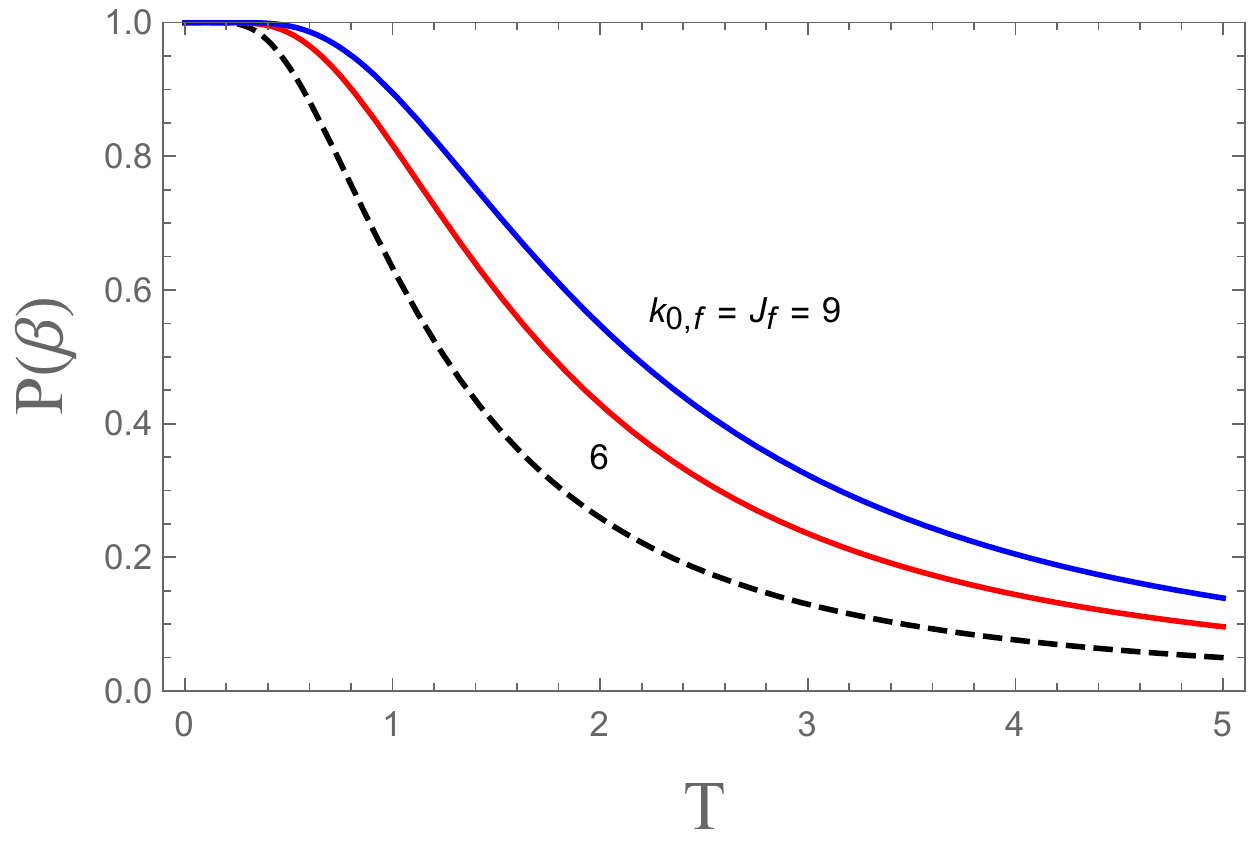} \hspace{0.5cm}
\includegraphics[height=5.0cm]{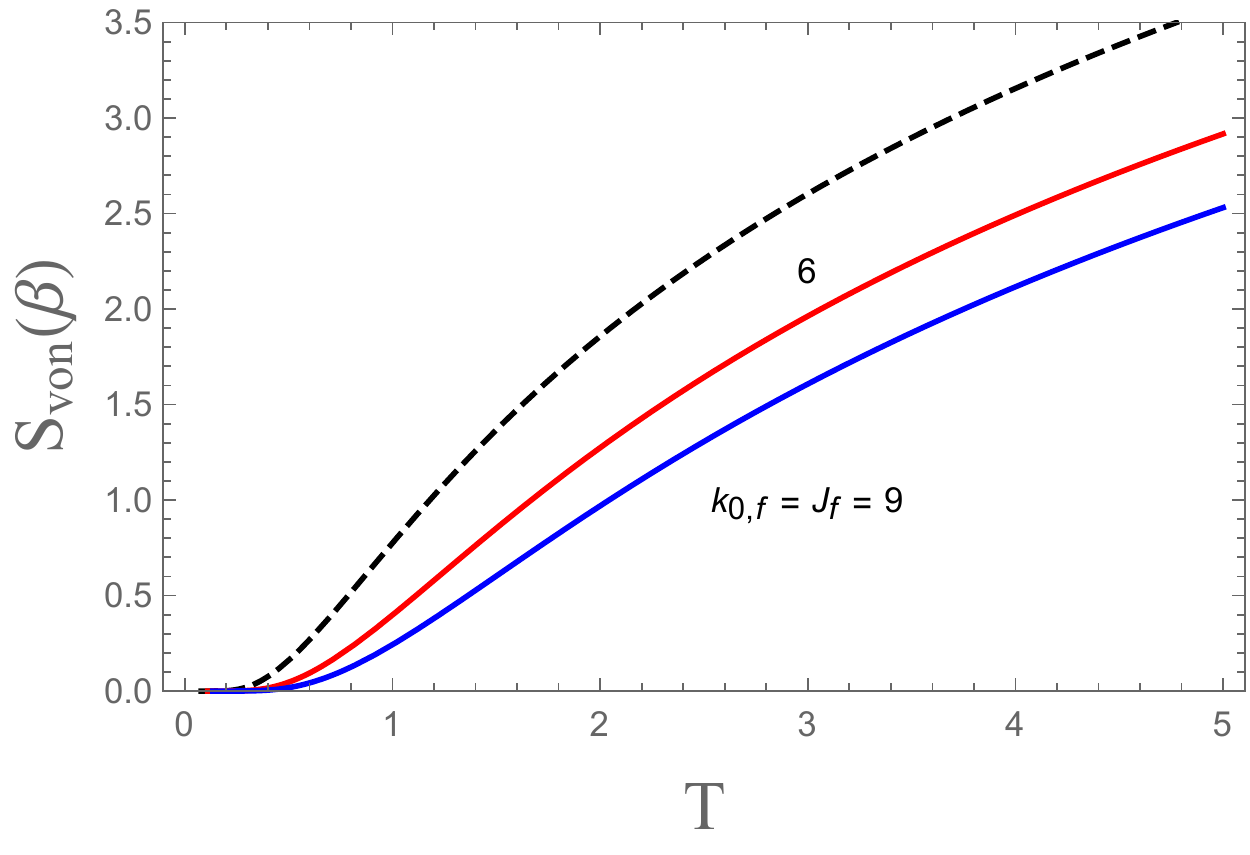}
\includegraphics[height=5.0cm]{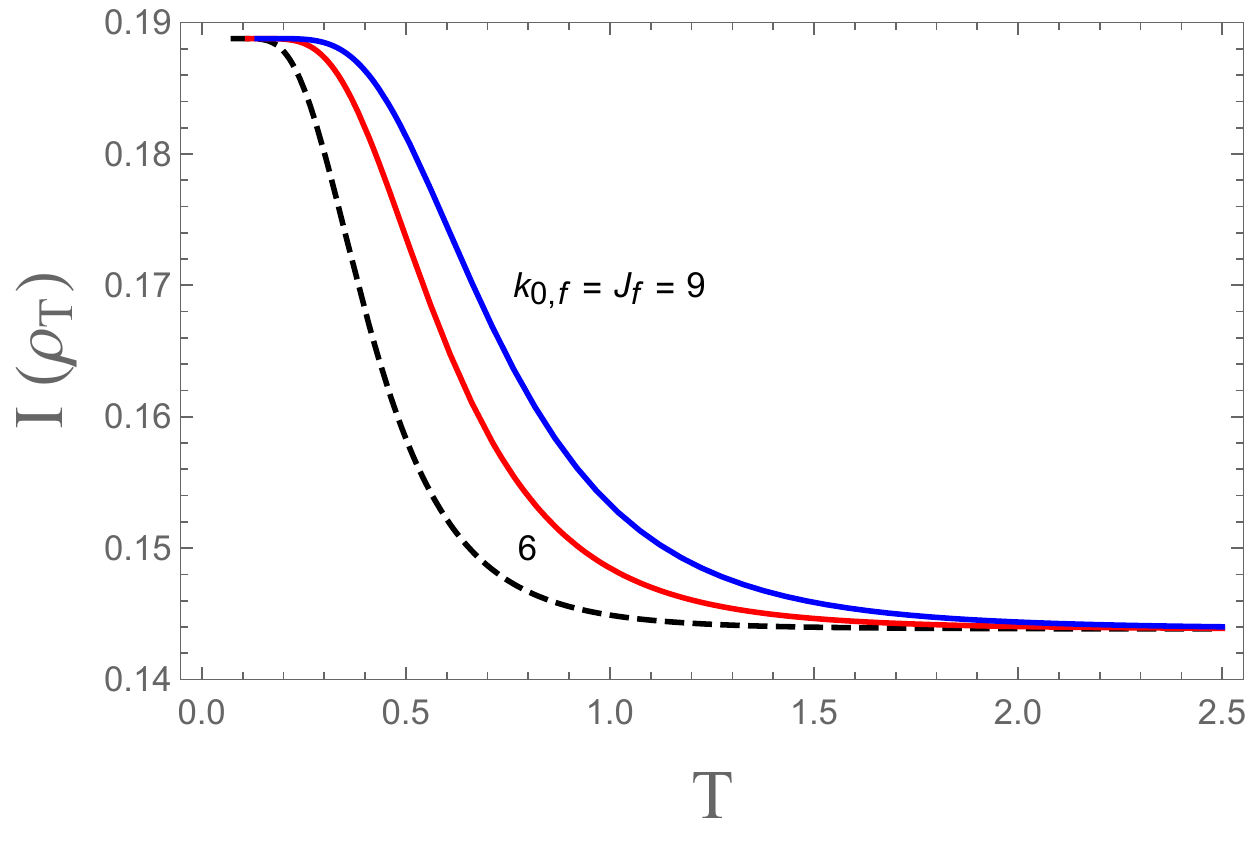}

\caption[fig2]{(Color online) The $T$-dependence of (a) $P(\beta)$ (b) $S_{von}$, and (c) $I (\rho_T)$ when $k_{0,f} = J_f = 6$ (red line) and  $k_{0,f} = J_f = 9$  (blue line) when $k_{0,i} = J_i = 3$.
The black dashed line corresponds to constant frequencies $k_{0, i} = J_i = k_{0,f} = J_f = 3$. 
 }
\end{center}
\end{figure}
%%%%%%%%%%%%%%%%%%%%%%%%%%%%%%%%%%%%%%%%%%%%%%%%%%%%%%%%%%%

The temperature-dependence of the purity function is plotted in Fig. 2 (a) when $k_{0,f} = J_f = 6$ (red line) and  $k_{0,f} = J_f = 9$  (blue line). The $k_{0,i}$ 
and $J_i$ are fixed as  $k_{0,i} = J_i = 3$. The black dashed line corresponds to constant frequencies $k_0 = J = 3$. As expected $\rho_T$ becomes more and more mixed with increasing temperature. 
Fig. 2(a) also show that $\rho_T$ is less mixed when  $|k_{0, f} - k_{0, i}|$ and $|J_f - J_i|$ increase.

\subsection{ R\'{e}nyi and von Neumann entropies}

In order to solve the R\'{e}nyi and von Neumann entropies of $\rho_T$ we should solve the eigenvalue equation
\begin{equation}
\label{eigen3-1}
\int dx_1 dx_2 \rho_T [x_1', x_2':x_1, x_2:\beta] u_{mn} (x_1, x_2:\beta) = p_{mn} (\beta) u_{mn} (x_1', x_2':\beta).
\end{equation}
Eq. (\ref{eigen3-1}) is solved in appendix A and the eigenvalue $p_{mn} (\beta)$ is 
\begin{equation}
\label{eigen3-2}
p_{mn} (\beta) = (1 - \xi_1) \xi_1^m (1 - \xi_2) \xi_2^n
\end{equation}
where
\begin{equation}
\label{eigen3-3}
\xi_1 = \frac{\sqrt{a_{1,+}} - \sqrt{a_{1,-}}}{\sqrt{a_{1,+}} + \sqrt{a_{1,-}}},      \hspace{1.0cm}
\xi_2 = \frac{\sqrt{a_{2,+}} - \sqrt{a_{2,-}}}{\sqrt{a_{2,+}} + \sqrt{a_{2,-}}}.
\end{equation}
In terms of $\xi_1$ and $\xi_2$ the purity function $P(\beta)$ in Eq. (\ref{purity3-2}) can be written as 
\begin{equation}
\label{purity3-4}
P (\beta) = \frac{1 - \xi_1}{1 + \xi_1} \frac{1 - \xi_2}{1 + \xi_2}.
\end{equation}
Then, the R\'{e}nyi and von Neumann entropies of $\rho_T$ reduce to 
\begin{equation}
\label{von3-1}
S_{\alpha} = S_{1, \alpha} + S_{2, \alpha}    \hspace{1.0cm} S_{von} = S_{1,von} + S_{2,von}
\end{equation}
where 
\begin{equation}
\label{von3-2}
S_{j,\alpha} = \frac{1}{1 - \alpha} \ln \frac{(1 - \xi_j)^{\alpha}}{1 - \xi_j},      \hspace{1.0cm}
S_{j,von} = -\ln (1 - \xi_j) - \frac{\xi_j}{1 - \xi_j} \ln \xi_j
\end{equation}
with $j=1,2$. 

One can show also that the normalized eigenfunction $u_{mn} (x_1, x_2:\beta)$ is 
\begin{equation}
\label{eigen3-4}
u_{mn}(x_1,x_2:\beta) =  \left( \frac{1}{{\cal C}_{1,m}} H_m (\sqrt{\epsilon_1} y_1) e^{-\mu_1 y_1^2} \right)
                                      \left( \frac{1}{{\cal C}_{2,n}} H_n (\sqrt{\epsilon_2} y_2) e^{-\mu_2 y_2^2} \right)
\end{equation}
where
\begin{eqnarray}
\label{eigen3-5}
&&\epsilon_1 = \sqrt{a_{1,+} a_{1,-}}   \hspace{4.2cm}  \epsilon_2 = \sqrt{a_{2,+} a_{2,-}}                  \\    \nonumber
&&\mu_1 = \frac{1}{2} \left[ \epsilon_1 + (\alpha_1 - \alpha_2) - (\alpha_3 - \alpha_4) \right],      \hspace{.5cm}
    \mu_2 = \frac{1}{2} \left[ \epsilon_2 + (\alpha_1 - \alpha_2) + (\alpha_3 - \alpha_4) \right]           \\    \nonumber
&& {\cal C}_{1,m}^2 = \frac{1}{\sqrt{2 \mu_1}} \sum_{k=0}^m 2^{2m-k} \left(\frac{\epsilon_1}{2\mu_1} - 1 \right)^{m-k}
\frac{\Gamma^2 (m+1) \Gamma (m- k + 1/2)}{\Gamma(k+1) \Gamma^2 (m-k+1)}                     \\    \nonumber
&& {\cal C}_{2,n}^2 = \frac{1}{\sqrt{2 \mu_2}} \sum_{k=0}^n 2^{2n-k} \left(\frac{\epsilon_2}{2\mu_2} - 1 \right)^{n-k}
\frac{\Gamma^2 (n+1) \Gamma (n- k + 1/2)}{\Gamma(k+1) \Gamma^2 (n-k+1)} .
\end{eqnarray}
For the case of constant frequencies $\alpha_1 - \alpha_2 = \alpha_3 - \alpha_4 = 0$, which results in $\mu_j = \frac{\epsilon_j}{2}$. In this case the sum in 
${\cal C}_{1.m}^2$ or ${\cal C}_{2,n}^2$ is nonzero only when $k=m$ or $k=n$, and this fact yields well-known quantities 
${\cal C}_{1,m}^{-1} = \frac{1}{\sqrt{2^m m!}} \left( \frac{\epsilon_1}{\pi} \right)^{1/4}$ and 
${\cal C}_{2,n}^{-1} = \frac{1}{\sqrt{2^n n!}} \left( \frac{\epsilon_2}{\pi} \right)^{1/4}$. Thus, the spectral decomposition of $\rho_T$ can be written as 
\begin{equation}
\label{spectral3-1}
\rho_T [x_1',x_2':x_1,x_1:\beta] = \sum_{m,n} p_{mn} (\beta) u_{mn} (x_1',x_2':\beta) u_{mn}^* (x_1,x_2:\beta).
\end{equation}

The temperature-dependence of the von Neumann entropy is plotted in Fig. 2 (b) when $k_{0,f} = J_f = 6$ (red line) and  $k_{0,f} = J_f = 9$  (blue line). The  $k_{0,i}$ and $J_i$ are fixed as  $k_{0,i} = J_i = 3$. The black dashed line corresponds to constant frequencies $k_0 = J = 3$. As expected $\rho_T$ becomes more and more mixed with increasing temperature. Fig. 2(b) also show that $\rho_T$ is less entangled when  $|k_{0, f} - k_{0, i}|$ and $|J_f - J_i|$ increase as purity function exhibits.  

\subsection{mutual information}

From $\rho_T$ in Eq. (\ref{thermal2-3}) one can derive the substates $\rho_{T,A} = \mbox{tr}_B \rho_T$ and $\rho_{T,B} = \mbox{tr}_A \rho_T$
by performing partial trace appropriately. Then, the substates become
\begin{equation}
\label{mutual3-1}
\rho_{T,A}[x', x:\beta] = \rho_{T,B} [x',x:\beta] = \sqrt{\frac{a_{1,-} a_{2,-}}{\pi (\alpha_1 + \alpha_2 - 2 \alpha_5)}} e^{-B_1 x^2 - B_2 x'^2 + 2 B_3 x x'}
\end{equation}
where 
\begin{eqnarray}
\label{mutual3-2}
&&B_1 = \frac{\alpha_2 (\alpha_1 + \alpha_2 - 2 \alpha_5) - (\alpha_4 + \alpha_6)^2}{\alpha_1 + \alpha_2 - 2 \alpha_5},    \hspace{0.2cm}
B_2 = \frac{\alpha_1 (\alpha_1 + \alpha_2 - 2 \alpha_5) - (\alpha_3 + \alpha_6)^2}{\alpha_1 + \alpha_2 - 2 \alpha_5}    \\    \nonumber
&&\hspace{2.5cm}B_3 = \frac{\alpha_5  (\alpha_1 + \alpha_2 - 2 \alpha_5) + (\alpha_3 + \alpha_6) (\alpha_4 + \alpha_6)}{\alpha_1 + \alpha_2 - 2 \alpha_5}.
\end{eqnarray}
It is not difficult to show that the eigenvalues of $\rho_{T,A}$ or $\rho_{T,B}$ are $(1 - \zeta) \zeta^n$, where 
\begin{equation}
\label{mutual3-3}
\zeta = \frac{2 B_3}{(B_1 + B_2) + \nu}
\end{equation}
with $\nu = \sqrt{(B_1 + B_2)^2 - 4 B_3^2}$.
Using the eigenvalues the R\'{e}nyi and von Neumann entropies of $\rho_{T,A}$ and $\rho_{T,B}$ can be obtained as 
\begin{equation}
\label{mutual3-4}
S_{A,\alpha} = S_{B,\alpha} = \frac{1}{1 - \alpha} \ln \frac{(1 - \zeta)^{\alpha}}{1 - \zeta^{\alpha}},    \hspace{.5cm}
S_{A,von} = S_{B,von} = - \ln (1 - \zeta) - \frac{\zeta}{1 - \zeta} \ln \zeta.
\end{equation}
Therefore, the mutual information of $\rho_T$ is given by 
\begin{equation}
\label{mutual3-5}
I (\rho_T) = S_{A,von} +  S_{B,von} - S_{von}.
\end{equation}

The temperature-dependence of the mutual information is plotted in Fig. 2 (c) when $k_{0,f} = J_f = 6$ (red line) and  $k_{0,f} = J_f = 9$  (blue line). The  $k_{0,i}$ and $J_i$ are fixed as  $k_{0,i} = J_i = 3$. The black dashed line corresponds to constant frequencies $k_0 = J = 3$. Like other quantities mutual information also decreases with 
increasing temperature. However, it does not completely vanish at $T = \infty$. Fig. 2 (c) shows that the mutual information seems to approach $0.144$ at the 
large temperature limit. This implies that the common information parties $A$ and $B$ share does not completely vanish even in the infinity temperature limit.

\section{Thermal Entanglement Phase Transition: Case of SQM}
Since the thermal state $\rho_T$ given in Eq. (\ref{thermal2-3}) is mixed state, its entanglement is in general defined via the convex-roof 
method\cite{benn96,uhlmann99-1};
\begin{equation}
\label{final-5}
{\cal E} (\rho_T) = \min \sum_j p_j {\cal E} (\psi_j),
\end{equation}
where minimum is taken over all possible pure state decompositions, i.e. $\rho_T = \sum_j p_j \ket{\psi_j} \bra {\psi_j}$, with $0 \leq p_j \leq 1$ 
and $\sum_j p_j = 1$.
The decomposition which yields minimum value is called the optimal decomposition. 
However, it seems to be highly difficult problem to derive the optimal decomposition in the continuous variable system.

Because of this difficulty, we will consider the negativity-like quantity\cite{vidal01} of $\rho_T$. Let $\sigma_T$ be a partial transpose of $\rho_T$, i.e.,
\begin{eqnarray}
\label{partial-T-1}
&&\sigma_T [x'_1,x'_2:x_1, x_2: \beta]                                                                                          
\equiv \rho_T [x_1, x'_2:x'_1,x_2:\beta]                                                                                                \\     \nonumber
&=& \frac{\sqrt{a_{1,-} a_{2,-}}}{\pi} 
\exp \bigg[ -\alpha_1 (x_1^2 + x_2'^2) - \alpha_2 (x_1'^2 + x_2^2) + 2 \alpha_3 x_1 x_2' + 2 \alpha_4 x_1' x_2     \\    \nonumber
&& \hspace{6.0cm} + 2 \alpha_5 (x_1 x_1' + x_2 x_2') + 2 \alpha_6 (x_1 x_2 + x_1' x_2')  \bigg].
\end{eqnarray}
Then, the negativity-like quantity ${\cal N} (\rho_T)$ is defined as 
\begin{equation}
\label{negat-1}
{\cal N} (\rho_T) = \sum_{m,n} |\Lambda_{mn}| - 1, 
\end{equation}
where $\Lambda_{mn}$ is eigenvalue of $\sigma_T$, i.e.,
\begin{equation}
\label{negat-2}
\int dx_1 dx_2 \sigma_T [x_1', x_2':x_1, x_2:\beta] f_{mn} (x_1, x_2) = \Lambda_{mn} (\beta) f_{mn} (x_1', x_2': \beta).
\end{equation}

One may wonder why the negativity-like quantity is introduced, because the PPT is known as necessary and sufficient criterion of 
separability for only $2 \times 2$ qubit-qubit and $2 \times 3$ qubit-qudit states\cite{peres96,horodecki96,horodecki97}. However, as Ref. \cite{duan-2000,simon-2000} have shown, PPT also provides 
a necessary and sufficient criterion of the separability for Gaussian continuous variable quantum states. Furthermore, the distillation protocols to maximally entangled state have been already suggested in Ref. \cite{duan-00-1,giedke-2000} in the Gaussian states. Thus, our negativity-like quantity is valid at least to determine whether the given Gaussian state is entangled or not. 
Since ${\cal N} (\rho_T)$ is proportional to ${\cal E} (\rho_T)$, ${\cal N} (\rho_T) = 0$ at the critical temperature $T= T_c$ of the TEPT 
if the external temperature induces the ESD phenomenon. 
Thus, if the eigenvalue equation (\ref{negat-2}) is solved, it is possible to compute $T_c$. 

As we will show in the following, however, it seems to be very difficult to solve Eq. (\ref{negat-2}) directly. In order to solve Eq. (\ref{negat-2}) we define
\begin{equation}
\label{negat-3}
f_{mn} (x_1, x_2: \beta) = e^{\frac{\alpha_1 - \alpha_2}{2} (x_1^2 - x_2^2)} g_{mn} (x_1, x_2, \beta).
\end{equation}
Then, Eq. (\ref{negat-2}) can be written as 
\begin{eqnarray}
\label{negat-4}
&&\frac{\sqrt{a_{1,-} a_{2,-}}}{\pi} \exp \left[ - \frac{\alpha_1 + \alpha_2}{2} \left( x_1'^2 + x_2'^2 \right) + 2 \alpha_6 x_1' x_2' \right]
                                                                                                                                                                                              \\    \nonumber
&&\times \int dx_1 dx_2 \exp \left[ - \frac{\alpha_1 + \alpha_2}{2} (x_1^2 + x_2^2) + 2 \alpha_6 x_1 x_2 + 
                                                       2 \left(  \begin{array}{cc}
                                                                     x_1',  &  x_2'
                                                                      \end{array}    \right)    \left(   \begin{array}{cc}
                                                                                       \alpha_5  &  \alpha_4  \\
                                                                                       \alpha_3  &  \alpha_5
                                                                                          \end{array}                \right)    \left(   \begin{array}{c}
                                                                                                                                                                x_1    \\
                                                                                                                                                                x_2       
                                                                                                                                                        \end{array}     \right)  \right]     \\   \nonumber
&&\hspace{6.0cm}  \times g_{mn} (x_1, x_2: \beta)      = \Lambda_{mn} (\beta) g_{mn} (x_1', x_2': \beta).
\end{eqnarray}
If one changes the variables as $y_1 = \frac{1}{\sqrt{2}} (x_1 + x_2)$ and $y_2 = \frac{1}{\sqrt{2}} (-x_1 + x_2)$, Eq. (\ref{negat-4}) reduces to 
\begin{eqnarray}
\label{negat-5}
&&\frac{\sqrt{a_{1,-} a_{2,-}}}{\pi} e^{-\mu_- y_1'^2 - \mu_+ y_2'^2}
\int dy_1 dy_2 \exp \left[ -\mu_- y_1^2 - \mu_+ y_2^2 + \left(     \begin{array}{cc}  y_1',  &  y_2'   \end{array} \right) A    
                                                                              \left(    \begin{array}{c}   y_1   \\   y_2   \end{array}   \right)    \right]    \\    \nonumber
&& \hspace{5.0cm} \times g_{mn} (y_1, y_2: \beta)  = \Lambda_{mn} (\beta)  g_{mn} (y_1', y_2': \beta)
\end{eqnarray}
where 
\begin{eqnarray}
\label{negat-6}
\mu_{\pm} = \frac{\alpha_1 + \alpha_2}{2} \pm \alpha_6,  \hspace{1.0cm}
A = \left(                     \begin{array}{cc}
                           2 \alpha_5 + (\alpha_3 + \alpha_4)   &   - (\alpha_3 - \alpha_4)     \\
                           \alpha_3 - \alpha_4     &     2 \alpha_5 - (\alpha_3 + \alpha_4)
                                   \end{array}                                               \right).
\end{eqnarray}
The difficulty arises because of the fact that $A$ is not symmetric matrix if $\alpha_3 \neq \alpha_4$. Due to this fact it seems to be impossible to 
factorize Eq. (\ref{negat-5}) into two single-party eigenvalue equations as we did in appendix A. 

However, Eq. (\ref{negat-5}) can be solved for the case of constant frequencies, i.e., $\omega_{1,i} = \omega_{1,f} \equiv \omega_1$ and 
$\omega_{2,i} = \omega_{2,f} \equiv \omega_2$, because in this case $\alpha_3$ is exactly equals to $\alpha_4$. Furthermore, in this case we get
\begin{eqnarray}
\label{negat-7}
&&\mu_+ = \frac{1}{4} \left[ \omega_1 \coth \frac{\omega_1 \beta}{2} + \omega_2 \tanh \frac{\omega_2 \beta}{2} \right],   \hspace{.5cm}
\mu_- = \frac{1}{4} \left[ \omega_1 \tanh \frac{\omega_1 \beta}{2} + \omega_2 \coth \frac{\omega_2 \beta}{2} \right]    \nonumber   \\
&& \hspace{2.0cm} \nu_+ \equiv \alpha_5 - \frac{\alpha_3 + \alpha_4}{2} = \frac{1}{4} \left[ \omega_1 \coth \frac{\omega_1 \beta}{2} - \omega_2 \tanh \frac{\omega_2 \beta}{2} \right]                                                                                                                                     \\    \nonumber
&& \hspace{2.0cm} \nu_- \equiv \alpha_5 + \frac{\alpha_3 + \alpha_4}{2} = -\frac{1}{4} \left[ \omega_1 \tanh \frac{\omega_1 \beta}{2} - \omega_2 \coth \frac{\omega_2 \beta}{2} \right].
\end{eqnarray}
Since $\alpha_3 = \alpha_4$ in this case, Eq. (\ref{negat-5}) is factorized into the following two single-party eigenvalue equations:
\begin{eqnarray}
\label{negat-8}
&&e^{-\mu_- y_1'^2} \int dy_1 e^{-\mu_- y_1^2 + 2 \nu_- y_1' y_1} g_{1,m} (y_1: \beta) = p_m (\beta) g_{1,m} (y_1':\beta)        \\     \nonumber
&&e^{-\mu_+ y_2'^2} \int dy_2 e^{-\mu_+ y_2^2 + 2 \nu_+ y_2' y_2} g_{2,n} (y_2: \beta) = q_n (\beta) g_{2,n} (y_2':\beta).
\end{eqnarray}
Then, the total eigenvalue $\Lambda_{mn}$ and the normalized eigenfunction $f_{mn} (x_1, x_2: \beta)$ are expressed as 
\begin{eqnarray}
\label{negat-9}
&&\Lambda_{mn} = \frac{1 }{\pi}\sqrt{\omega_1 \omega_2 \tanh \frac{\omega_1 \beta }{2} \tanh \frac{\omega_2 \beta}{2}} p_m (\beta) q_n (\beta)  
                                                                                                                                                                          \\  \nonumber
&& f_{mn} (x_1, x_2:\beta) = g_{1,m} (y_1:\beta) g_{2,n} (y_2:\beta),
\end{eqnarray}
where $g_{1,m} (y_1:\beta)$ and $g_{2,n} (y_2:\beta)$ are normalized eigenfunctions of Eq. (\ref{negat-8}). 
Solving Eq. (\ref{negat-8}) it is straightforward to show that the normalized eigenfunctions are 
\begin{eqnarray}
\label{negat-10}
&&g_{1,m} (y_1:\beta) = \frac{1}{\sqrt{2^m m!}} \left( \frac{\epsilon_1}{\pi} \right)^{1/4} H_m \left( \sqrt{\epsilon_1} y_1 \right) e^{-\frac{\epsilon_1}{2} y_1^2}                                                                                                                                                             \\    \nonumber
&&g_{2,n} (y_2:\beta) = \frac{1}{\sqrt{2^n n!}} \left( \frac{\epsilon_2}{\pi} \right)^{1/4} H_n \left( \sqrt{\epsilon_2} y_2 \right) e^{-\frac{\epsilon_2}{2} y_2^2},
\end{eqnarray}
where
\begin{eqnarray}
\label{negat-11}
&&\epsilon_1 = 2 \sqrt{\mu_-^2 - \nu_-^2} = \sqrt{\omega_1 \omega_2 \tanh \frac{\omega_1 \beta}{2} \coth \frac{\omega_2 \beta}{2}}
                                                                                                                                                                       \\    \nonumber
&&\epsilon_2 = 2 \sqrt{\mu_+^2 - \nu_+^2} = \sqrt{\omega_1 \omega_2 \coth \frac{\omega_1 \beta}{2} \tanh \frac{\omega_2 \beta}{2}}.
\end{eqnarray}
One can also show that the eigenvalue $\Lambda_{mn}$ is 
\begin{equation}
\label{negat-12}
\Lambda_{mn} = (1 - \zeta_1) (1 - \zeta_2) \zeta_1^m \zeta_2^n
\end{equation}
where
\begin{eqnarray}
\label{negat-13}
&&\zeta_1 = \frac{\nu_-}{\mu_- + \frac{\epsilon_1}{2}} = \frac{\sqrt{\mu_- + \nu_-} - \sqrt{\mu_- - \nu_-}}{\sqrt{\mu_- + \nu_-} + \sqrt{\mu_- - \nu_-}} = - \frac{\sqrt{\omega_1 \tanh \frac{\omega_1 \beta}{2}} - \sqrt{ \omega_2 \coth \frac{\omega_2 \beta}{2}}}
                             {\sqrt{\omega_1 \tanh \frac{\omega_1 \beta}{2}} + \sqrt{ \omega_2 \coth \frac{\omega_2 \beta}{2}}}      \\    \nonumber
&&\zeta_2 = \frac{\nu_+}{\mu_+ + \frac{\epsilon_2}{2}} = \frac{\sqrt{\mu_+ + \nu_+} - \sqrt{\mu_+ - \nu_+}}{\sqrt{\mu_+ + \nu_+} + \sqrt{\mu_+ - \nu_+}} =  \frac{\sqrt{\omega_1 \coth \frac{\omega_1 \beta}{2}} - \sqrt{ \omega_2 \tanh \frac{\omega_2 \beta}{2}}}
                             {\sqrt{\omega_1 \coth \frac{\omega_1 \beta}{2}} + \sqrt{ \omega_2 \tanh \frac{\omega_2 \beta}{2}}}.
\end{eqnarray}

%%%%%%%%%%%%%%%%%%%%%%%%%%%%%%%%%%%%%%%%%%%%%%%%%%%%%%%%%
\begin{figure}[ht!]
\begin{center}
\includegraphics[height=5.0cm]{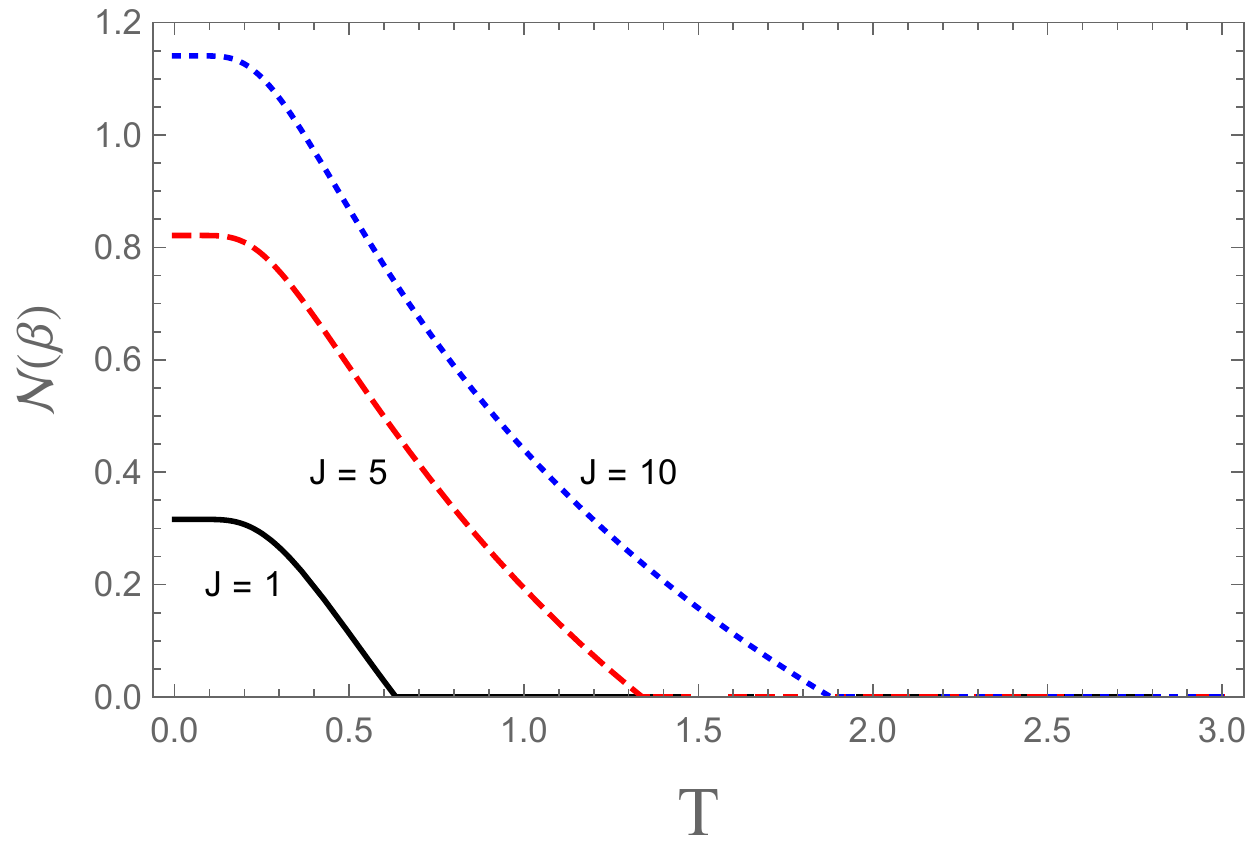} \hspace{0.5cm}
\includegraphics[height=5.0cm]{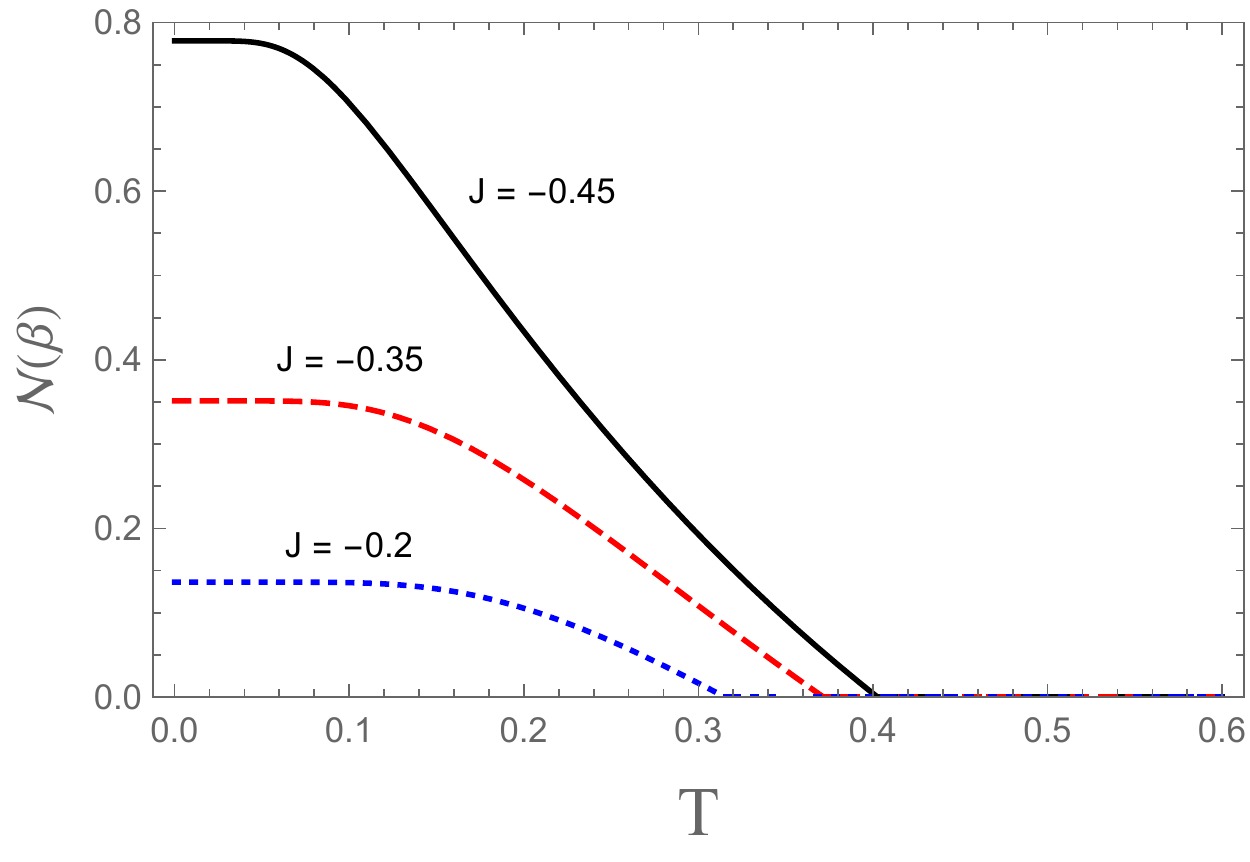}

\caption[fig3]{(Color online) The $T$-dependence of ${\cal N} (\beta)$  for (a) $J = 1$ (black line), $5$ (red dashed line), 
$10$ (blue dotted line) and (b) $J = -0.45$ (black line), $-0.35$ (red dashed line), $-0.2$ (blue dotted line) with fixing $k_0 = 1$. Both figures show 
${\cal N} (\beta)$ is zero at $T \geq T_c$. Since ${\cal N} (\beta)$ is proportional to entanglement of $\rho_T$, this fact implies that 
$\rho_T$ is entangled (or separable) state at $T < T_c$ (or $T \geq T_c$). The critical temperature temperature $T_c$ increases with increasing 
$|J|$.  }
\end{center}
\end{figure}
%%%%%%%%%%%%%%%%%%%%%%%%%%%%%%%%%%%%%%%%%%%%%%%%%%%%%%%%%%%

One can compute $\pm 1 - \zeta_1$ and $\pm 1 - \zeta_2$ explicitly, which result in $-1 < \zeta_1, \zeta_2 \leq 1$ for arbitrary temperature. Thus, 
it is easy to show $\sum_{m,n} \Lambda_{mn} (\beta) = 1$ as expected. Eq. (\ref{negat-1}) and Eq. (\ref{negat-12}) make ${\cal N} (\beta)$
to be 
\begin{equation}
\label{nega-2}
{\cal N} (\beta) = \frac{(1 - \zeta_1) (1 - \zeta_2)}{(1 - |\zeta_1|) (1 - |\zeta_2|)} - 1.
\end{equation}
The $T$-dependence of ${\cal N} (\beta)$ is plotted in Fig. 3 for (a) positive and (b) negative $J$ with fixing $k_0 = 1$. Both figures show 
${\cal N} (\beta)$ is zero at $T \geq T_c$. Similar results were obtained for general bosonic harmonic lattice systems\cite{peres96,plenio}.
Since ${\cal N} (\beta)$ is proportional to entanglement of $\rho_T$, this fact implies that 
$\rho_T$ is entangled (or separable) state at $T < T_c$ (or $T \geq T_c$). The critical temperature temperature $T_c$ increases with increasing 
$|J|$. 

From Eq. (\ref{nega-2}) it is evident that $\rho_T$ is separable when $\zeta_1 \geq 0$ and $\zeta_2 \geq 0$. Eq. (\ref{negat-13}) implies that 
this separability criteria can be rewritten in a form
\begin{equation}
\label{separable-1}
x \tanh x - y \coth y \leq 0,      \hspace{1.0cm}   x \coth x - y \tanh y \geq 0
\end{equation}
where $x = \omega_1 \beta / 2$ and $y = \omega_2 \beta / 2$. If $J \geq 0$, first equation of Eq. (\ref{separable-1}) is automatically satisfied. Hence, 
the second equation plays a role as a genuine separability criterion. If $J < 0$, first equation is true criterion. It is worthwhile noting that two equations
in Eq. (\ref{separable-1}) can be transformed into each other by interchanging $x$ and $y$. This fact implies that the region in $x$-$y$ plane, where the 
separable states reside, is symmetric with respect to $y = x$. 

%%%%%%%%%%%%%%%%%%%%%%%%%%%%%%%%%%%%%%%%%%%%%%%%%%%%%%%%%
\begin{figure}[ht!]
\begin{center}
\includegraphics[height=5.0cm]{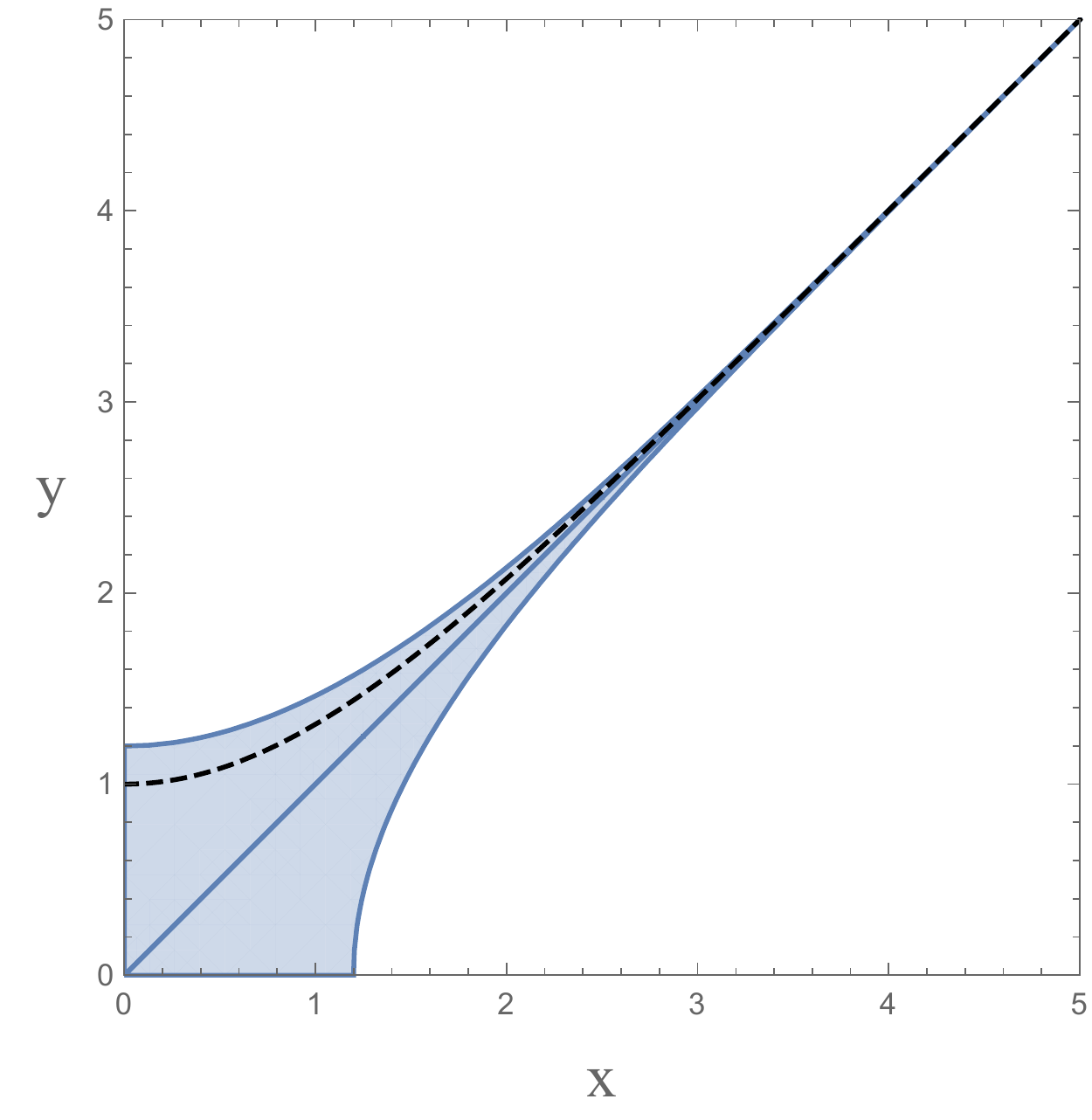} \hspace{0.5cm}
\includegraphics[height=5.0cm]{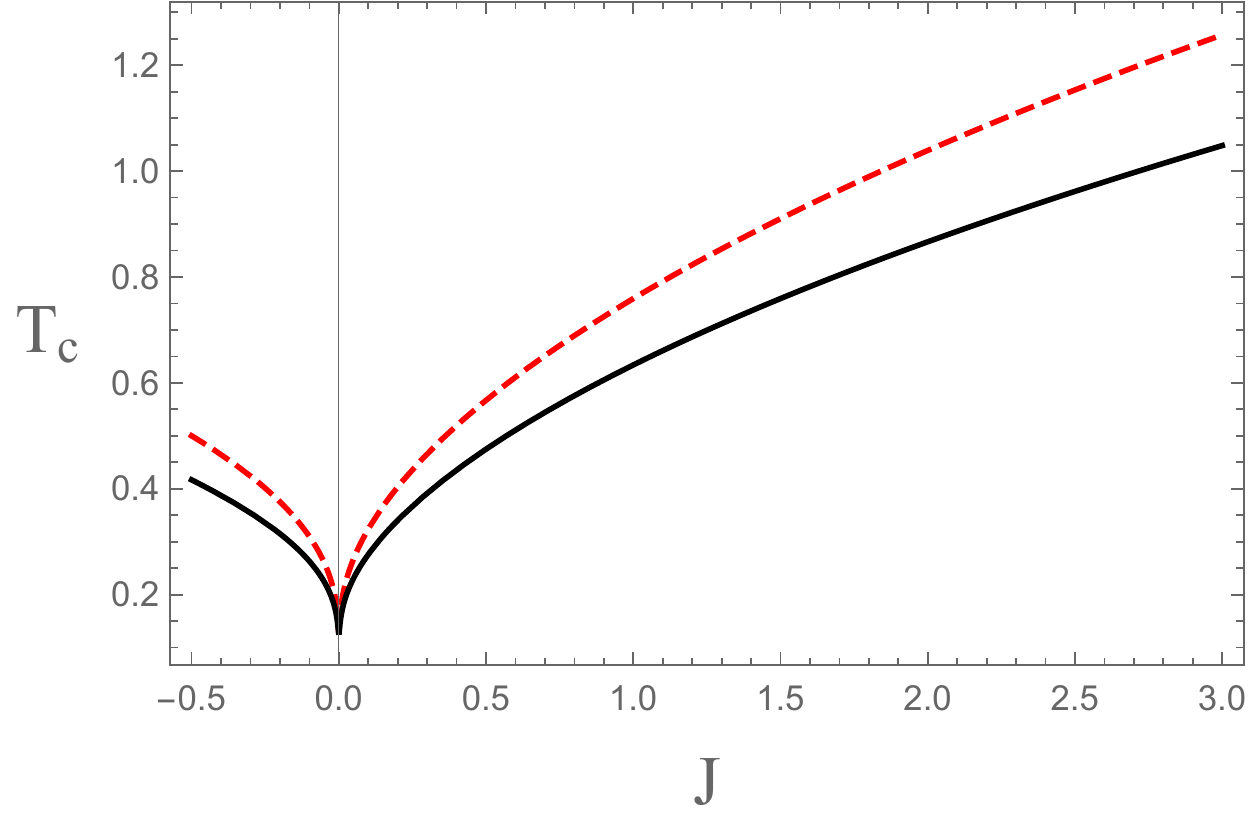}

\caption[fig4]{(Color online) (a) The shaded region  in $x$-$y$ plane is a region where the separable states of $\rho_T$ reside. The boundary contains an information on the critical 
temperature $T_c$. The black dashed line in the shaded region is $y = x \coth x$. This is used to compute $T_c$ approximately. 
(b)  The $J$-dependence of $T_c$ when $k_0 = 1$. The black solid line and red dashed line correspond to exact and approximate
$T_c$ respectively.}
\end{center}
\end{figure}
%%%%%%%%%%%%%%%%%%%%%%%%%%%%%%%%%%%%%%%%%%%%%%%%%%%%%%%%%%%

The shaded region in Fig. 4(a) is a region where the separable states of $\rho_T$ reside in $x$-$y$ plane. As expected, the region is symmetric with respect to $y = x$. 
It is shown that most separable states are accumulated in $0 \leq x, y \leq 1$. The boundary of the region contains an information about the critical
temperature $T_c$. The black dashed line in the region is $y = x \coth x$. Since this is very close to upper boundary, this can be used to compute $T_c$
approximately. 

Let the upper boundary of Fig. 4(a) be expressed by $y_c = x_c g (x_c)$, where $x_c$ and $y_c$ are $x$ and $y$ at $T = T_c$. Then the low boundary should
be $x_c = y_c g (y_c)$. The function $g(z)$ can be derived numerically by using Eq. (\ref{separable-1}) after changing the inequality into equality. Then, 
$T_c$ can be computed by 
\begin{equation}
\label{critical-1}
T_c = \frac{\omega_{min}}{2 g^{-1} \left( \frac{\omega_{max}}{\omega_{min}} \right)}
\end{equation}
where $\omega_{min} = \min (\omega_1, \omega_2)$ and  $\omega_{max} = \max (\omega_1, \omega_2)$. If one uses $g(z) \approx \coth z$,
the critical temperature is approximately 
\begin{equation}
\label{critical-2}
T_c \approx \frac{\omega_{min}}{\ln \left(\frac{\omega_{max} + \omega_{min}}{\omega_{max} - \omega_{min}} \right)}.
\end{equation}
In Fig. 4(b) the $J$-dependence of $T_c$ is plotted when $k_0 = 1$. The black solid line and red dashed line correspond to Eq. (\ref{critical-1}) and 
Eq. (\ref{critical-2}) respectively. It is shown that $T_c$ increases with increasing $|J|$ as expected from Fig. 3. 

%%%%%%%%%%%%%%%%%%%%%%%%%%%%%%%%%%%%%%%%%%%%%%%%%%%%%%%%%
\begin{figure}[ht!]
\begin{center}
\includegraphics[height=5.0cm]{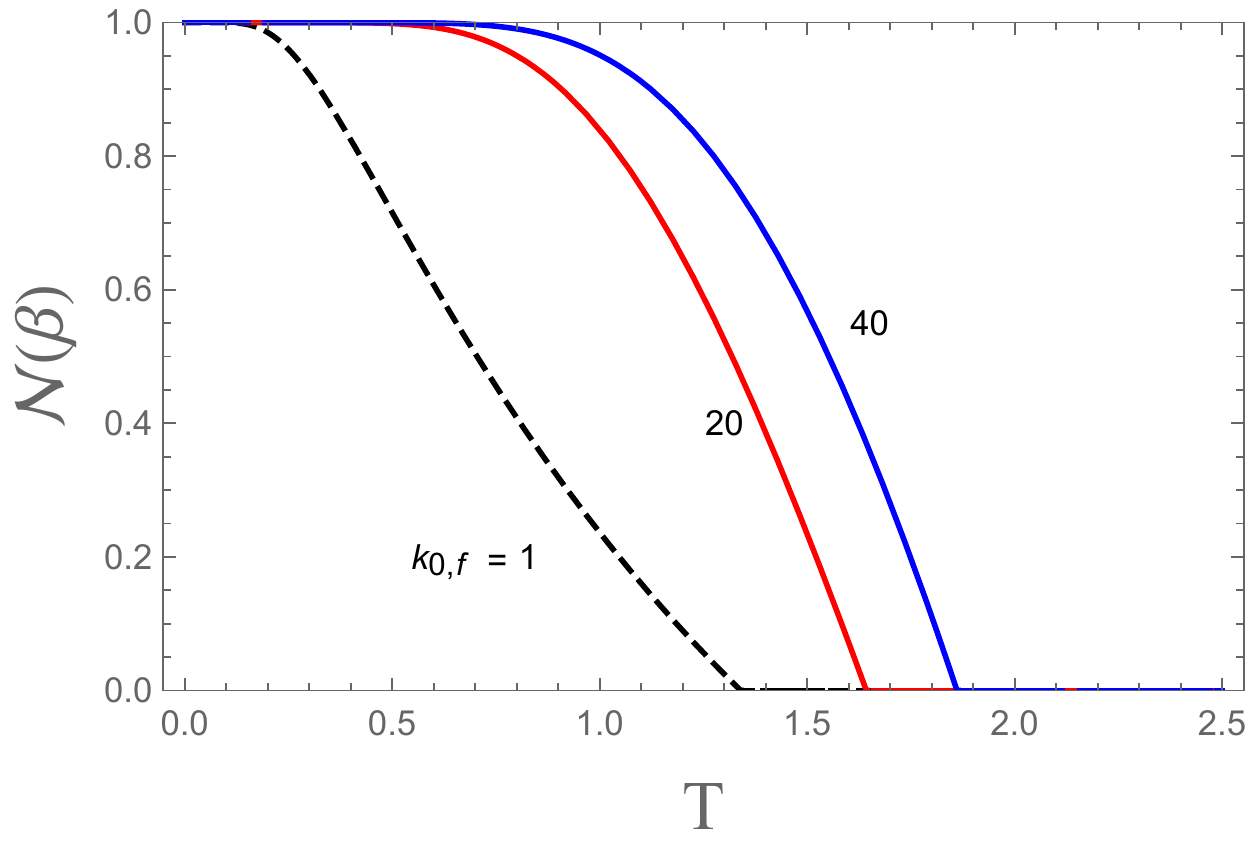} \hspace{0.5cm}
\includegraphics[height=5.0cm]{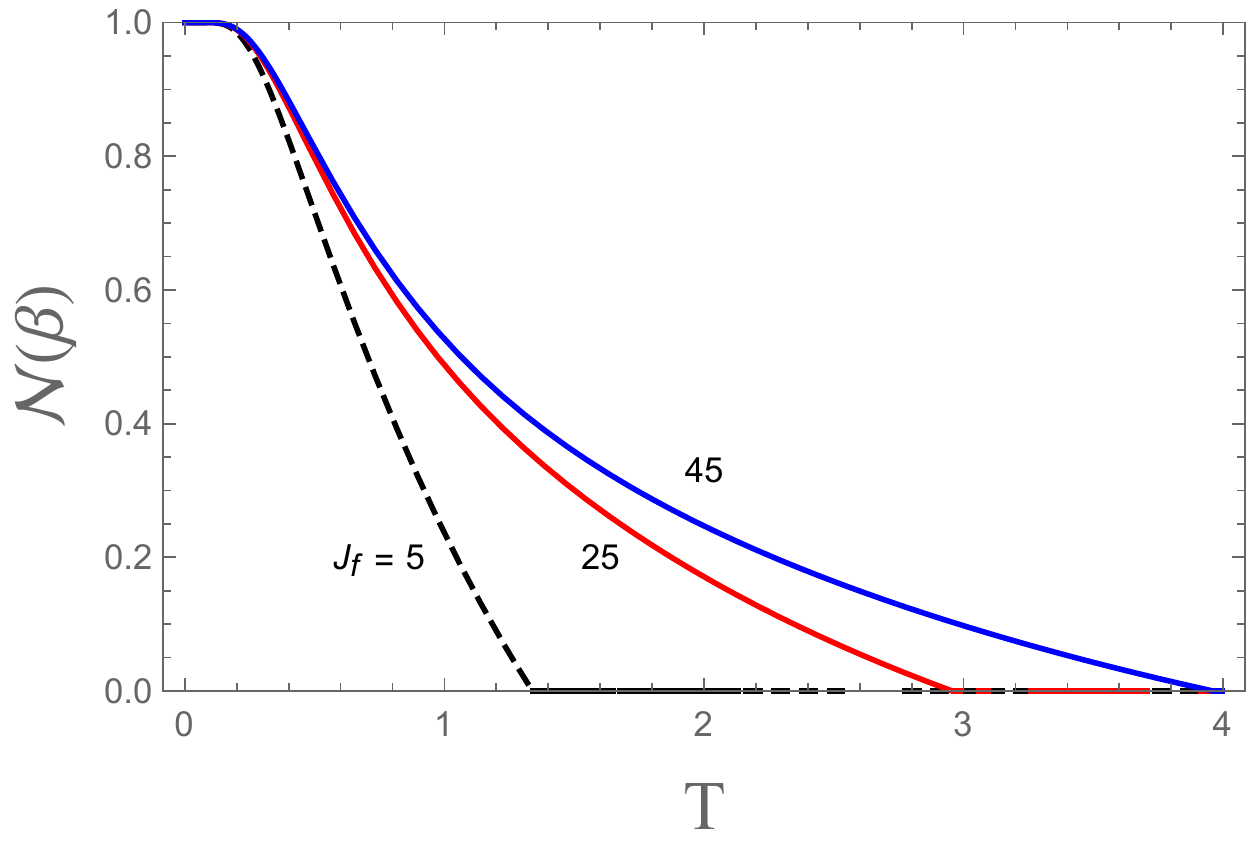}

\caption[fig4]{(Color online) (a) The temperature dependence of ${\cal N} (\beta) / {\cal N} (\infty)$ when $k_{0,f} = 1$ (black dashed line), 
$k_{0,f} = 20$ (red line) and $k_{0,f} = 40$ (blue line) with fixing $k_{0,i} = 1$ and $J_i = J_f = 5$. 
(b) The temperature dependence of ${\cal N} (\beta) / {\cal N} (\infty)$ when $J_f = 5$ (black dashed line), $J_f = 25$ (red line), and $J_f = 45$ (blue line)
with fixing $k_{0,i} = k_{0,f} = 1$ and $J_i = 5$.  Both figures exhibit that the critical temperature $T_c$ increases with increasing $|k_{0,f} - k_{0,i}|$ and 
$|J_f - J_i|$. }
\end{center}
\end{figure}
%%%%%%%%%%%%%%%%%%%%%%%%%%%%%%%%%%%%%%%%%%%%%%%%%%%%%%%%%%%

As we commented earlier, for the case of SQM it seems to be highly difficult problem to solve the eigenvalue equation (\ref{negat-2}) directly. However, 
we can conjecture the eigenvalue $\Lambda_{mn} (\beta)$ without deriving the eigenfunction $f_{mn} [x_1, x_2:\beta]$ as follows. Since 
$\sum_{m,n} \Lambda_{mn} (\beta) = 1$, $\Lambda_{mn} (\beta)$ might be represented as Eq. (\ref{negat-12}). If this is right, we can compute 
$\zeta_1$ and $\zeta_2$ by making use of  the R\'{e}nyi entropy. 
If the eigenvalue is represented as Eq. (\ref{negat-12}),  the R\'{e}nyi entropy of $\sigma_T$
can be written as 
\begin{equation}
\label{negat-18}
S_{\alpha} [\sigma_T] \equiv \frac{1}{1 - \alpha} \ln \mbox{tr} \left[ (\sigma_T)^{\alpha} \right] 
= \frac{1}{1 - \alpha} \left[ \ln \frac{(1 - \zeta_1)^{\alpha}}{1 - \zeta_1^{\alpha}} + \ln \frac{(1 - \zeta_2)^{\alpha}}{1 - \zeta_2^{\alpha}}\right].
\end{equation}
Putting $\alpha=2$ or $3$ in Eq. (\ref{negat-18}), it is possible to derive
\begin{equation}
\label{negat-19}
\frac{(1 - \zeta_1) (1 - \zeta_2)}{(1 + \zeta_1) (1 + \zeta_2)} = \beta_1,    \hspace{1.0cm}
\frac{(1 - \zeta_1)^2 (1 - \zeta_2)^2}{(1 + \zeta_1 + \zeta_1^2) (1 + \zeta_2 + \zeta_2^2)} = \beta_2
\end{equation}
where 
\begin{equation}
\label{negat-20}
\beta_1 \equiv \mbox{tr} \sigma_T^2 = \sqrt{\frac{X_1}{X_2}},     \hspace{1.0cm}
\beta_2 \equiv \mbox{tr} \sigma_T^3 = \frac{4 X_1}{X_1 + 3 X_2 - 12 (\alpha_5^2 - \alpha_3 \alpha_4)}
\end{equation}
with 
\begin{equation}
\label{negat-21}
X_1 = (\alpha_1 + \alpha_2 - 2 \alpha_5)^2 - (\alpha_3 + \alpha_4 + 2 \alpha_6)^2,    \hspace{.5cm}
X_2 = (\alpha_1 + \alpha_2 + 2 \alpha_5)^2 - (\alpha_3 + \alpha_4 - 2 \alpha_6)^2.
\end{equation}
Solving Eq. (\ref{negat-19}), we get 
\begin{eqnarray}
\label{negat-22}
&&u = \zeta_1 + \zeta_2 = \frac{1 - \beta_1}{2 (4 \beta_1^2 - \beta_1^2 \beta_2 - 3 \beta_2)} \left[ -3 \beta_2 (1 + \beta_1) + \sqrt{3 \beta_2 \left[ 16 \beta_1^2 - \beta_2 (3 - \beta_1)^2 \right]} \right]                                                     \\       \nonumber
&& v = \zeta_1 \zeta_2 = -1 +  \frac{1 + \beta_1}{2 (4 \beta_1^2 - \beta_1^2 \beta_2 - 3 \beta_2)} \left[ -3 \beta_2 (1 + \beta_1) + \sqrt{3 \beta_2 \left[ 16 \beta_1^2 - \beta_2 (3 - \beta_1)^2 \right]} \right]. 
\end{eqnarray}
Thus, $\zeta_1$ and $\zeta_2$ for the case of SQM become
\begin{equation}
\label{negat-23}
\zeta_1 = \frac{u + \sqrt{u^2 - 4 v}}{2},     \hspace{1.0cm}  \zeta_2 = \frac{u - \sqrt{u^2 - 4 v}}{2}.
\end{equation}
At the case of constant frequency $\beta_1$ and $\beta_2$ reduce to 
\begin{equation}
\label{negat-24}
\beta_1 = \frac{\epsilon_1 \epsilon_2}{4 (\mu_+ + \nu_+) (\mu_- + \nu_-)},        \hspace{1.0cm}
\beta_2 = \frac{4 (\mu_+ - \nu_+) (\mu_- - \nu_-)}{(2 \mu_+ + \nu_+) (2 \mu_- + \nu_-)}.
\end{equation}
Using Eq. (\ref{negat-24}) and after tedious calculation, one can show that $\zeta_1$ and $\zeta_2$ in Eq. (\ref{negat-23}) exactly coincide with those in 
Eq. (\ref{negat-13}) when $\omega_{1,i} = \omega_{1,f} = \omega_1$ and $\omega_{2,i} = \omega_{2,f} = \omega_2$. Then, the negativity-like 
quantity can be written in a form
\begin{equation}
\label{negat-25}
{\cal N} (\beta) = \frac{1 - u + v}{1 + |v| - (|\zeta_1| + |\zeta_2|)} - 1.
\end{equation}

The temperature dependence of ${\cal N} (\beta) / {\cal N} (\infty)$ is plotted in Fig. 5. In Fig. 5(a) we choose $k_{0,f} = 1$ (black dashed line), 
$k_{0,f} = 20$ (red line), and $k_{0,f} = 40$ (blue line) when $k_{0,i} = 1$ and $J_i = J_f = 5$. As this figure exhibits, the critical temperature $T_c$ increases with 
increasing $|k_{0,f} - k_{0,i}|$. In Fig. 5(b) we choose $J_f = 5$ (black dashed line), $J_f = 25$ (red line), and $J_f = 45$ (blue line) when $k_{0,i} = k_{0,f} = 1$
and $J_i = 5$. This figure also shows that $T_c$ increases with increasing $|J_f - J_i|$.

\section{Conclusions}
In this paper we derive explicitly the thermal state of the two coupled harmonic oscillator system when the spring and coupling constants are 
arbitrarily time-dependent. In particular, we focus on the SQM model (see Eq. (\ref{instant2-1}) and Eq. (\ref{instant2-2})). In this model we compute purity 
function,  R\'{e}nyi and von Neumann entropies, and mutual information analytically and examine their temperature-dependence. We also discuss on the TEPT by making use of the negativity-like quantity. Our calculation shows that the critical temperature $T_c$ increases with 
increasing the difference between the initial and final frequencies. In this way we can use the SQM model to protect the entanglement against the external temperature by introducing a large difference of frequencies, i.e. $|\omega_f - \omega_i| \gg 1$. 

There are several issues related to our paper. Since the SQM model we consider involves a discontinuity at $t=0$, it is unrealistic in some sense. In order to 
escape this fact we can introduce the time-dependence of frequencies as a form $\omega = \omega_i + (\omega_f - \omega_i) \sin \Omega t$.
Then, we have to solve the Ermakov equation numerically. In this case the critical temperature $T_c$ might be dependent on $\Omega$ and 
$|\omega_f - \omega_i|$. Then, it may be possible to protect the entanglement in the thermal bath by adjusting  $\Omega$ and $|\omega_f - \omega_i|$
appropriately.

In this paper we introduce the negativity-like quantity to examine the thermal entanglement, because we do not know how to derive the 
optimal decomposition of Eq. (\ref{final-5}). Recently, the upper and lower bounds of entanglement of formation (EoF) are examined for arbitrary two-mode
Gaussian state\cite{ralph19}. It seems to be of interest to examine the TEPT with EoF.

%{\bf Acknowledgement}:
%On April 16, 2014 the ferry Sewol has sunk into the South Sea of Korea. Due to this disaster 304 people died and, 9 of them are still missing. We would like to dedicate this paper to all victims of this accident.
%This research was supported by the Basic Science Research Program through the National Research Foundation of Korea(NRF) funded by the Ministry of Education, Science and Technology(2011-0011971).
%This work was supported by the Kyungnam University Foundation Grant, 2018.

\newpage 

\begin{appendix}{\centerline{\bf Appendix A}}

\setcounter{equation}{0}
\renewcommand{\theequation}{A.\arabic{equation}}

In this section we examine the eigenvalue equation of  the following bipartite Gaussian state:
\begin{eqnarray}
\label{type2}
&& \rho_2 [x_1', x_2': x_1, x_2]     
 = A \exp \bigg[ -a_1 (x_1'^2 + x_2'^2) - a_2 (x_1^2 + x_2^2) + 2 b_1 x_1' x_2' + 2 b_2 x_1 x_2     \\    \nonumber
&&  \hspace{7.0cm} + 2 c (x_1 x_1' + x_2 x_2') + 2 f (x_1 x_2' + x_2 x_1')  \bigg]
\end{eqnarray}
where $A =  \sqrt{(a_1 + a_2 - 2c)^2 - (b_1 + b_2 + 2 f)^2} / \pi$. If $a_1 = \alpha_1$, $a_2 = \alpha_2$, $b_1 = \alpha_3$, $b_2 = \alpha_4$, $c = \alpha_5$, 
and $f = \alpha_6$, $\rho_2$ is exactly the same with the thermal state $\rho_T$ given in Eq. (\ref{thermal2-3}). 
Now let us consider the eigenvalue equation
\begin{equation}
\label{eigen4-1}
\int dx_1 dx_2 \rho_2 [x_1', x_2': x_1, x_2] f_{mn} (x_1, x_2) = \lambda_{mn} f_{mn} (x_1', x_2').
\end{equation}

First we change the variables as 
\begin{equation}
\label{change4-1}
y_1 = \frac{1}{\sqrt{2}} (x_1 + x_2),   \hspace{2.0cm}  y_2 = \frac{1}{\sqrt{2}} (x_1 - x_2).
\end{equation}
Then Eq. (\ref{eigen4-1}) is simplified as 
\begin{eqnarray}
\label{eigen4-2}
&& A e^{-(a_1 - b_1) y_1'^2 - (a_1 + b_1) y_2'^2} 
\int dy_1 dy_2 e^{- (a_2 - b_2) y_1^2 - (a_2 + b_2) y_2^2 + 2 (c + f) y_1' y_1 + 2 (c - f) y_2' y_2} f_{mn} (y_1, y_2)   \nonumber   \\
&&  \hspace{8.0cm}= \lambda_{mn} f(y_1', y_2').
\end{eqnarray}

Now, we define 
\begin{equation}
\label{define4-1}
f_{mn} (y_1, y_2) = g_m (y_1) h_n (y_2).
\end{equation}
Then, Eq. (\ref{eigen4-2}) is solved if one solves the following two single-party eigenvalue equations:
\begin{eqnarray}
\label{eigen4-3}
&& e^{-(a_1 - b_1) y_1'^2} \int dy_1 e^{-(a_2 - b_2) y_1^2 + 2 (c + f) y_1' y_1} g_m (y_1) = p_m g_m (y_1')       \\     \nonumber
&& e^{-(a_1 + b_1) y_2'^2} \int dy_2 e^{-(a_2 + b_2) y_2^2 + 2 (c - f) y_2' y_2} h_n (y_2) = q_n h_n (y_2').
\end{eqnarray}
The eigenvalue of Eq. (\ref{eigen4-1}) can be computed as $\lambda_{mn} = A p_m q_n$. 

By making use of Eq. (\ref{von-2}) and Eq. (\ref{von-3}) one can show $\lambda_{mn} = (1 - \xi_1) \xi_1^m (1 - \xi_2) \xi_2^n$, where
\begin{eqnarray}
\label{eigen4-4}
&&\xi_1 = \frac{2 (c + f)}{(a_1 + a_2 - b_1 - b_2) + \epsilon_1}                                                                          \\   \nonumber
&&\hspace{.5cm}= \frac{\sqrt{(a_1 + a_2 - b_1 - b_2) + 2 (c+f)} - \sqrt{(a_1 + a_2 - b_1 - b_2) - 2 (c+f)}}
                                                                                                        {\sqrt{(a_1 + a_2 - b_1 - b_2) + 2 (c+f)} + \sqrt{(a_1 + a_2 - b_1 - b_2) - 2 (c+f)}}
                                                                                                                                                                        \\   \nonumber
&&\xi_2 = \frac{2 (c - f)}{(a_1 + a_2 + b_1 + b_2) + \epsilon_2}                                                                          \\   \nonumber
&&\hspace{.5cm}= \frac{\sqrt{(a_1 + a_2 + b_1 + b_2) + 2 (c-f)} - \sqrt{(a_1 + a_2 + b_1 + b_2) - 2 (c-f)}}
                                      {\sqrt{(a_1 + a_2 + b_1 + b_2) + 2 (c-f)} + \sqrt{(a_1 + a_2 + b_1 + b_2) - 2 (c-f)}}
\end{eqnarray}
with
\begin{equation}
\label{eigen4-5}
\epsilon_1 = \sqrt{(a_1 + a_2 - b_1 - b_2)^2 -4 (c + f)^2},     \hspace{1.0cm} \epsilon_2 = \sqrt{(a_1 + a_2 + b_1 + b_2)^2 -4 (c - f)^2}.
\end{equation}

We can also use Eq. (\ref{von-2}) and Eq. (\ref{von-3}) to derive the normalized eigenfunction, whose explicit expression is 
\begin{equation}
\label{eigen4-6}
f_{mn} (x_1, x_2) = \left(\frac{1}{{\cal C}_{1,m}} H_m (\sqrt{\epsilon_1} y_1) e^{-\frac{\alpha_1}{2} y_1^2} \right) 
                            \left(\frac{1}{{\cal C}_{2,n}} H_n (\sqrt{\epsilon_2} y_2) e^{-\frac{\alpha_2}{2} y_2^2} \right)  
\end{equation}  
where 
\begin{equation}
\label{eigen4-7}
\alpha_1 = \epsilon_1 + (a_1 - a_2) - (b_1 - b_2),    \hspace{1.0cm}  \alpha_2 = \epsilon_2 + (a_1 - a_2) + (b_1 - b_2)
\end{equation}    
and the normalization constants ${\cal C}_{1,m}$ and ${\cal C}_{2,n}$ are   
\begin{eqnarray}
\label{eigen4-8}
&&{\cal C}_{1,m}^2 = \frac{1}{\sqrt{\alpha_1}} \sum_{k=0}^m 2^{2m - k} \left( \frac{\epsilon_1}{\alpha_1} - 1 \right)^{m-k} 
    \frac{\Gamma^2 (m+1) \Gamma (m - k + 1/2)}{\Gamma (k + 1) \Gamma^2 (m-k+1)}                                \\    \nonumber
&&{\cal C}_{2,n}^2 = \frac{1}{\sqrt{\alpha_2}} \sum_{k=0}^n 2^{2n - k} \left( \frac{\epsilon_2}{\alpha_2} - 1 \right)^{n-k} 
    \frac{\Gamma^2 (n+1) \Gamma (n - k + 1/2)}{\Gamma (k + 1) \Gamma^2 (n-k+1)}. 
\end{eqnarray} 
Thus, the spectral decomposition of $\rho_2$ is 
\begin{equation}
\label{sepctral4-1}
\rho_2 [x_1', x_2':x_1, x_2] = \sum_{m,n} \lambda_{mn} f_{mn} (x_1', x_2') f_{mn}^* (x_1, x_2),
\end{equation}
where $\lambda_{mn}$ and $f_{mn}$ are given in Eq. (\ref{eigen4-4}) and Eq. (\ref{eigen4-6}) respectively.

\end{appendix}

\end{document}